\documentclass[12pt]{article}
\usepackage{amssymb}
\usepackage{graphics}
\usepackage{epsfig}
\usepackage{a4wide}

\begin{document}
\newcommand{\ppbwphj}{$p\bar{p}/pp \to W^{+} H^0j+X$ }
\newcommand{\ppbwhj}{$p\bar{p}/pp \to W^{-} H^0j+X$ }
\newcommand{\pbwhj}{$p\bar{p} \to W^{\pm} H^0j+X$ }
\newcommand{\ppwhjm}{$pp \to W^{-} H^0j+X$ }
\newcommand{\ppwhjp}{$pp \to W^{+} H^0j+X$ }

\title{ Next-to-leading order QCD predictions for the hadronic $WH$+jet production  }
\author{ Su Ji-Juan, Ma Wen-Gan, Zhang Ren-You, and Guo Lei   \\
{\small Department of Modern Physics, University of Science and Technology  }  \\
{\small of China (USTC), Hefei, Anhui 230026, People's Republic of China}  }

\date{}
\maketitle \vskip 15mm
\begin{abstract}
We calculate the next-to-leading order(NLO) QCD corrections to the
$WH^0$ production in association with a jet at hadron colliders.
We study the impacts of the complete NLO QCD radiative corrections
to the integrated cross sections, the scale dependence of the
cross sections, and the differential cross sections ($\frac{d
\sigma}{d\cos\theta}$, $\frac{d \sigma}{dp_T}$) of the final $W$-,
Higgs boson and jet. We find that the corrections significantly
modify the physical observables, and reduce the scale uncertainty
of the leading-order cross section. Our results show that by applying the
inclusive scheme with $p_{T,j}^{cut}=20~GeV$ and taking
$m_H=120~GeV$, $\mu=\mu_0\equiv\frac{1}{2}(m_W+m_H)$, the K-factor
is $1.15$ for the process $p\bar p \to W^{\pm}H^0j+X$ at the
Tevatron, while the K-factors for the processes $pp \to W^-H^0j+X$
and $pp \to W^+H^0j+X$ at the LHC are $1.12$ and $1.08$
respectively. We conclude that to understand the hadronic
associated $WH^0$ production, it is necessary to study the NLO QCD
corrections to the $WH^0j$ production process which is part of the
inclusive $WH^0$ production.
\end{abstract}

\vskip 5mm


\vfill \eject

\baselineskip=0.32in

\renewcommand{\theequation}{\arabic{section}.\arabic{equation}}
\renewcommand{\thesection}{\arabic{section}.}
\newcommand{\nb}{\nonumber}

\newcommand{\Dir}{\kern -6.4pt\Big{/}}
\newcommand{\Dirin}{\kern -10.4pt\Big{/}\kern 4.4pt}
\newcommand{\DDir}{\kern -7.6pt\Big{/}}
\newcommand{\DGir}{\kern -6.0pt\Big{/}}

\makeatletter      
\@addtoreset{equation}{section}
\makeatother       

\section{Introduction}
\par
In the standard model (SM) the Higgs mechanism explains the mass
generation, and is believed to be responsible for the breaking of
the electroweak (EW) symmetry\cite{1,2}. To discover the Higgs boson
and investigate thoroughly the mechanism of EW symmetry breaking are
the main physics motivations for the future high energy colliders.
At the Fermilab Tevatron, Higgs boson production associated with the $W$
or $Z^0$ boson, is the most important discovery channel for the SM
Higgs boson with light mass($m_H<135~GeV$)\cite{3,4}. At the CERN
LHC there are a few Higgs boson production mechanisms which lead to an
observable cross section. Each of them makes use of the preference
of couplings of the SM Higgs and massive gauge bosons or top
quarks\cite{5}. Recently, J.M. Butterworth, {\it et al.} concluded
that the subset techniques at the LHC have the potential to transform
the high-$p_T$ $WH^0$, $Z^0H^0$($H^0 \to b\bar b$) channel into one
of the best channels in finding a low mass SM Higgs and obtaining
the unique information on the coupling of the Higgs boson separately
to $W$ and $Z^0$ bosons\cite{lhchiggs}.

\par
At the TeV energy scale hadron colliders, the experimental
environment is extremely complicated. The produced signal and
background reactions normally involve multiparticles in the final
state. A good understanding of these reactions is very necessary
for studying the hadronic collider physics. It requires
sufficiently precise predictions for the new physics signals and
their backgrounds with multiple final particles which cannot
entirely be separated in experimental data. Therefore, high-order
predictions for these reactions are very useful. In fact, when we
measure experimentally the inclusive $WH^0$ production signal, it
includes any number of additional jets unless stated otherwise. In
this sense the $WH^0$+jet production is part of the inclusive
$WH^0$ production, and theoretically $WH^0$+jet at the next-to-leading order(NLO) QCD is
part of the $WH^0$ production process at the NNLO QCD. Recently,
the calculations of the QCD ${\cal O}(\alpha_s)$ and electroweak
${\cal O}(\alpha_{ew})$ corrections to the Higgs production
processes $p\bar{p}/pp \to WH^0/Z^0H^0+X$ at the Tevatron and the
LHC were presented in Refs. \cite{QCDc,Ewc},
respectively. The NNLO QCD corrections to the SM Higgs boson
production processes in association with the vector boson at hadron
colliders have been calculated in Ref.\cite{nnlo}.

\par
In this work we present precise calculations for the process
$p\bar{p}/pp \to WH^0j+X$ up to the QCD NLO at the Tevatron and
the LHC. The paper is organized as follows: We describe the
technical details of the related leading-order(LO) and NLO QCD calculations in
Secs. 2 and 3, respectively. In Sec.4 we give some numerical
results and discussions about the NLO QCD corrections. Finally a
short summary is given.

\par
\section{LO cross sections }
\par
\begin{figure*}
\begin{center}
\includegraphics[scale=1.0]{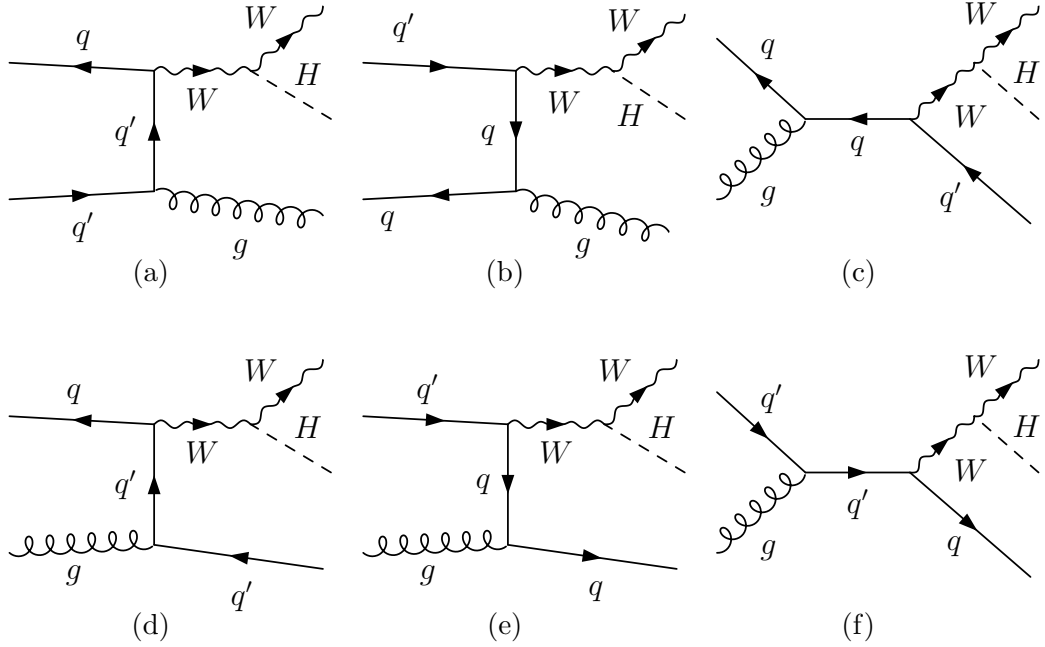}
\caption{\label{fig1} The generic LO Feynman diagrams for the
partonic processes $\bar{q} q'(g) \to W^{-}H^{0}g(\bar{q}')$ and
$q'g\to W^{-}H^{0} q$. Figures 1 (a) and (b) are the LO diagrams for
the partonic process $\bar q q' \to W^{-}H^0 g$, (c) and (d) for
the $\bar q g \to W^{-}H^0 \bar{q}'$, (e) and (f) for the $q' g
\to W^{-}H^0 q$, $(q=u,c;~q'=d,s)$.  }
\end{center}
\end{figure*}

At the partonic level the cross section for the $W^+H^0j$
production process in the SM should be the same as for the
$W^-H^0j$ production process due to the CP-conservation. We
present the LO calculations for the related partonic $W^-H^0j$
production processes in this section. We calculate the
$p\bar{p}/pp  \to W^{\pm}H^0j+X$ processes by neglecting u-, d-,
c-, s-, b-quark mass($m_u = m_d = m_c = m_s = m_b = 0$), and the
quark mixing between the third generation and other two
generations(i.e., $V_{ub} = V_{cb} = V_{td} = V_{ts} = 0$). In our
LO calculation we do not consider the partonic processes with
an incoming (anti)bottom-quark due to the heavy (anti)bottom-quark
suppression in parton distribution functions (PDFs) in the proton and
antiproton. Then the following partonic processes are involved in
our LO calculations.
\begin{eqnarray}
\label{process1} &&  \bar{q}(p_{1})+q'(p_{2})\to
W^{-}(p_{3})+H^0(p_{4})+g(p_{5}),  \\
\label{process2} &&  \bar{q}(p_{1})+g(p_{2})\to
W^{-}(p_{3})+H^0(p_{4})+\bar{q}'(p_{5}),  \\
\label{process3}&& q'(p_{1})+g(p_{2})\to
W^{-}(p_{3})+H^0(p_{4})+q(p_{5}),
\end{eqnarray}
where $q=u,c$; $q'=d,s$. $p_{i}(i=1,...,5)~$ represent the
four-momenta of the incoming partons and the outgoing $W^{-}$,
$H^0$ and jet, respectively. There are six LO Feynman diagrams for
all those partonic processes of the $W^{-} H^0 j$ production shown
in Fig.\ref{fig1}. There Figs.1(a,b), Figs.1(c,d) and Figs.1(e,f)
are the LO diagrams for the partonic process $\bar{q} q' \to W^-
H^0 g$, $\bar{q} g \to W^- H^0 \bar{q}'$ and $q' g \to W^- H^0 q$,
respectively.

\par
All partonic processes for $W^{-}H^0j$ production at hadron
colliders are related to the amplitude $0 \to W^{-}H^0q \bar{q}'g$
by crossing symmetry. The expressions of the LO cross sections for
the partonic processes $\bar{q} q' \to W^{-}H^0 g$, $\bar{q} g \to
W^{-}H^0 \bar{q}'$ and $q'g \to W^{-}H^0 q$ can be written in the
form as,
\begin{eqnarray}
\label{sigma_qqgg} \hat{\sigma}^{kl}_{LO}(\hat{s})&=& \frac{1}{2
\hat{s}} \int \overline{\sum}|{\cal M}_{LO}^{kl} |^2
d\Omega_{3}^{kl},~~(kl=\bar{q} q',\bar{q} g,q'g,~(q=u,c;~q'=d,s)),
\end{eqnarray}
where the summation is taken over the spins and colors of final
states, and the bar over the summation recalls averaging over the
spin and color degrees of freedom of initial partons. $d
\Omega_3^{kl}$ is the three-body phase-space element for the $kl \to
W^-H^0j$ channel expressed as
\begin{eqnarray}\label{PSelement}
d \Omega_3^{kl} = (2\pi)^4  \delta^4 (p_1+p_2-\sum_{i = 3}^5 p_i)
\prod_{j=3}^5 \frac{d^3 p_j}{(2\pi)^3 2E_j}.
\end{eqnarray}

\par
In Eq.(\ref{sigma_qqgg}) ${\cal M}_{LO}^{kl}$ is the amplitude of
the tree-level diagrams for anyone of the partonic processes
(\ref{process1})-(\ref{process3}). $\hat{s}$ is the partonic
center-of-mass energy squared. It is obvious that the LO cross
section $\hat{\sigma}_{LO}^{kl}$ is IR divergent when we integrate
the Feynman amplitude squared $| {\cal M}_{LO}^{kl} |^2$ over the
full three-body final state phase-space. The divergence arises from
the integration over the phase-space region where the final gluon is
soft or the final gluon/light-quark jet becomes collinear to one of
the initial partons. To avoid these IR singularities and obtain an
IR-safe result, we should take a transverse momentum cut for final
jet.

\par
The LO total cross sections for $p\bar{p}(pp) \to W^{-} H^0j+X$ can
be expressed as
\begin{eqnarray}\label{sigma_PP}
&& \sigma_{LO}(AB(p\bar{p},pp)\to W^{-}H^0j+X)=   \nonumber \\
&&
\sum_{kl=\bar{u}d,\bar{u}s,\bar{u}g,dg}^{\bar{c}d,\bar{c}s,\bar{c}g,sg}
\int dx_A dx_B \left[ G_{k/A}(x_A,\mu_f)
G_{l/B}(x_B,\mu_f)\hat{\sigma}^{kl}_{LO}(x_{A}x_{B}s,\mu_f)+(A\leftrightarrow
B)\right].
\end{eqnarray}
There $\mu_f$ is the factorization energy scale; $x_{A}$ and
$x_{B}$ describe the fractions of partons $k$ and $l$ in hadrons
(proton or antiproton) A and B respectively, with the definitions
of
\begin{eqnarray}\label{1}
x_{A}=\frac{p_{1}}{P_{A}},~~x_{B}=\frac{p_{2}}{P_{B}},
\end{eqnarray}
where $P_{A}$ and $P_{B}$ are the four-momenta of the incoming
hadrons A and B. $G_{i/H}$ ($i = u, d, c, s, g$, $H = p, \bar{p}$)
represent the PDFs of parton $i$ in hadron $H$. Analogous to 
Eq.(\ref{1}), we can obtain the expression for
$\sigma_{LO}(AB(p\bar{p},pp)\to W^{+}H^0j+X)$. For the LO
calculation we use the CTEQ6L1 PDFs\cite{cteq}.

\vskip 10mm
\section{ NLO QCD corrections }
\par
\subsection{Virtual corrections}
\par
In our numerical calculations we find if we take the nondiagonal
Cabibbo-Kobayashi-Maskawa(CKM) matrix, the contributions to the LO cross section for
$p\bar{p}(pp) \to W^{\pm} H^0j+X$ from the partonic processes
involving the coupling between the $W^{\pm}$-boson and quarks in two
different generations(i.e., $W^{+}-\bar{u}-s$ ($W^{-}-u-\bar{s}$)
and/or $W^{+}-\bar{c}-d$ ($W^{-}-c-\bar{d}$) couplings) are less
than $1.2\%$ and $5\%$ at the Tevatron and the LHC, respectively.
Therefore, it is reasonable to consider only the NLO QCD
corrections of partonic processes involving the coupling of the
$W^{\pm}$-boson with quarks in the same generation in our NLO
calculations.

\par
The virtual QCD corrections are evaluated in the t'Hooft-Feynman
gauge. We adopt the dimensional regularization (DR) scheme to
regulate the UV and IR divergences and the modified minimal
subtraction$(\overline{MS})$ scheme to renormalize the relevant
fields. The one-loop diagrams are essentially obtained from the
tree-level diagrams of related partonic processes of $W^{-}H^0j$
production, and modify the LO cross sections for partonic
processes (\ref{process1})-(\ref{process3}). These corrections are
induced by self-energy, vertex, box(4-point) and counterterm
diagrams. The one-loop level Feynman diagrams and corresponding
amplitudes are generated by using the FeynArts3.4 package\cite{fey}.
The amplitudes which involve UV and IR singularities, are further
analytically simplified by the modified FormCalc
programs\cite{formloop}. The final amplitudes are translated into
Fortran codes with the UV and IR ``$\epsilon \times$ N-point
integral'' terms remained unprocessed. The output amplitudes are
further numerically evaluated by using our developed Fortran
subroutines for calculating N-point integrals and extracting the
remaining finite $\epsilon \frac{1}{\epsilon}$ terms. In these
Fortran codes the IR singularities are separated from the
IR-finite remainder by adopting the expressions for the IR
singularity in N-point integrals($N\geq3$) in terms of 3-point
integrals\cite{Beenakk}. The whole reduction of tensor integrals
to the lower-rank tensors and further to the scalar integral is done
with the help of the LoopTools library \cite{formloop,Passarino},
and the FF package \cite{van}. The dimensionally regularized soft
or collinear singular 3- and 4-point integrals had to be added to
this library. The virtual corrections to the partonic processes
$kl \to W^{-}H^0j$ can be expressed as
\begin{eqnarray}\label{VirtualCS}
 d\hat{\sigma}_V^{kl}(\hat{s})
    = \frac{1}{2\hat{s}}\overline{\sum}2 Re
    \left({\cal M}_{LO}^{kl\dag}{\cal M}_V^{kl}
    \right) d \Omega_{3}^{kl},~~(kl=\bar{q} q',\bar{q} g,q'g),
\end{eqnarray}
where we use the same notations as in Eq.(\ref{sigma_qqgg});
${\cal M}_V^{kl}$ represents the renormalized one-loop amplitude
for the $kl$ annihilation partonic process. In both the LO and NLO
calculations for the $W^{-}H^0j$ production at the Tevatron and
the LHC we should involve all the contributions of the partonic
processes $kl \to W^{-}H^0j$ where $kl=\bar{u}d,~\bar{u}s$,
$\bar{c}d,~\bar{c}s$, $\bar{u}g,~\bar{c}g$, $dg,~sg$.

\par
There exist both ultraviolet(UV) and soft/collinear infrared(IR)
singularities in the loop corrections to the partonic process $kl
\to W^{-}H^0j$, but the total NLO QCD amplitude of the subprocess is
UV finite after performing the renormalization procedure.
Nevertheless, it still contains soft/collinear IR singularities.
The soft/collinear IR singularities can be cancelled by adding the
contributions of the real gluon/light-(anti)quark emission
partonic processes, and redefining the parton distribution
functions at the NLO.

\par
\subsection{Real gluon and light-(anti)quark emission corrections }
\par
The real gluon and light-(anti)quark emission partonic processes are
obtained from the matrix elements of $0 \to W^{-}H^0gg\bar{q}'q$ and
$0\to W^-H^0q\bar{q'}q''\bar{q}''$ by all possible crossings of
(anti)quarks($q$, $q'$ and $q''$) and gluons into the initial state. The
relevant real correction partonic processes can be grouped as: (1)
$gg \to W^{-}H^0\bar{q}' q$, (2) $\bar{q}q'\to W^{-}H^0gg$, (3)
$g\bar{q} \to W^{-}H^0g\bar{q}'$, (4) $gq' \to W^{-}H^0gq$, (5)
$\bar{q}q'\to W^{-}H^0q''\bar{q}''$, (6) $\bar{q}\bar{q}'' \to
W^{-}H^0\bar{q}'\bar{q}''$, (7) $q'' \bar{q} \to
W^{-}H^0\bar{q}'q''$, (8) $q'' \bar{q}'' \to W^{-}H^0q\bar{q}'$, (9)
$q'\bar{q}'' \to W^{-}H^0 q\bar{q}''$, (10) $q'q'' \to W^{-}H^0
qq''$. There the quark notations represent $q=u,c$, $q'=d,s$ and
$q''=u,d,c,s,b$, respectively. Since the (anti)bottom PDF in the
(anti)proton is heavily suppressed with respect to the other light
quarks, we neglect the real emission partonic processes which
involve the (anti)bottom quark in initial states. The real
gluon/light-(anti)quark emission partonic channels (1)-(10) at
tree-level give the origins of soft and collinear IR singularities.
After the summation of the virtual corrections with all the real
parton emission corrections, the numerical result is soft IR-safe,
while there still exists remained collinear divergence. But it will
be totally IR safe when we include the contributions from the
collinear counterterms of the PDFs.

\par
\begin{figure*}
\begin{center}
\includegraphics[scale=1.0]{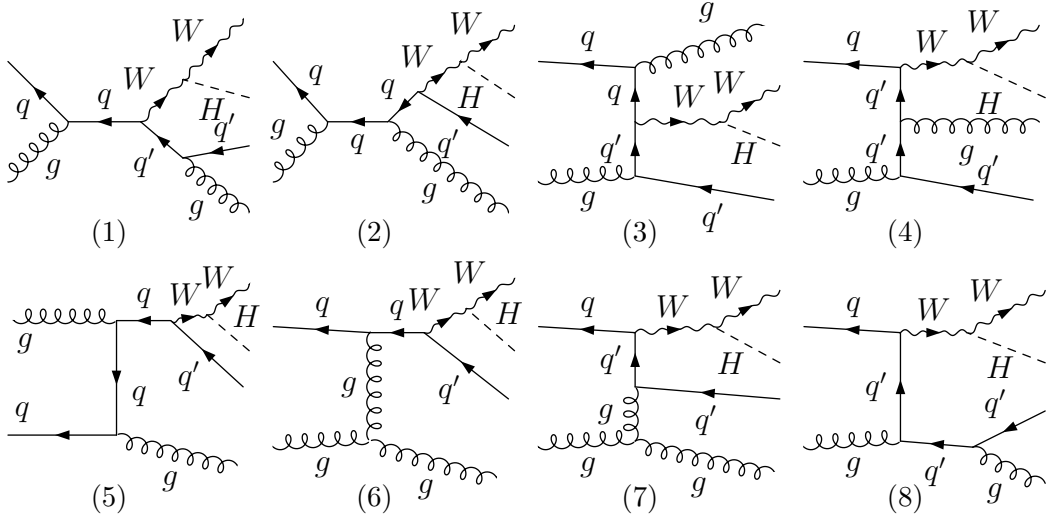}
\caption{\label{fig2} The tree-level Feynman diagrams for the real
emission process $\bar{q}g \to W^{-}H^0\bar{q}'g$. }
\end{center}
\end{figure*}

The IR singularities of the real parton emission subprocesses can be
isolated by adopting the two cutoff phase-space slicing (TCPSS)
method\cite{19}. We take the $\bar{q}(p_1)g(p_2)$ $\to
W^{-}(p_3)H^0(p_4)$ $\bar{q}'(p_5)g(p_6)$ ($q=u,c$, $q'=d,s$) as an
example and show how to deal with the calculation of the real
emission process. This partonic process contains eight LO Feynman
diagrams which are depicted in Fig.\ref{fig2}. We can find from
Fig.\ref{fig2} that the tree-level real emission subprocess
$\bar{q}g \to W^{-}H^0\bar{q}'g$ involves both the soft and
collinear singularities due to the gluon/antiquark($\bar{q},
\bar{q}'$) splitting in this initial or final state. The IR
singularities in the partonic process are isolated by applying the
TCPSS method. An arbitrary small soft cutoff $\delta_{s}$ is
introduced to separate the $2 \to 4$ phase-space into two regions,
$E_{6}\leq\delta_{s}\sqrt{\hat{s}}/2$(soft gluon region) and
$E_{6}>\delta_{s}\sqrt{\hat{s}}/2$ (hard gluon region). Another
cutoff $\delta_{c}$ is used to decompose the hard region into a hard
collinear(HC) region and hard noncollinear ($\overline{HC}$) region
to isolate the remaining collinear singularity from the soft IR-safe
hard region. The criterion for separating the HC region is described
as below: The region for real gluon/light-quark emission with
$\hat{s}_{16}(\hat{s}_{25}$, $\hat{s}_{26}$, $\hat{s}_{56})<
\delta_{c}\hat{s}$ (where $\hat{s}_{ij}=(p_{i}+p_{j})^{2}$) is
called the HC region. Otherwise it is called the $\overline{HC}$ region.
Then the cross section for the real emission partonic process
$\bar{q} g \to W^- H^0 \bar{q}' g$ can be written as
\begin{equation}
\label{sigmaR} \hat{\sigma}_{R}(\bar{q}g\to
W^{-}H^0\bar{q'}g)=\hat{\sigma}^{S}+\hat{\sigma}^{H}
=\hat{\sigma}^{S}+\hat{\sigma}^{HC}+\hat{\sigma}^{\overline{HC}}.
\end{equation}

\par
\subsection{NLO corrected cross sections  }
\par
The full NLO QCD corrected hadronic cross section for the
$W^{-}H^0j$ production at hadron colliders can be written as:
\begin{eqnarray}
&& \sigma_{NLO}(p\bar{p}/pp \to W^{-}H^0j+X) = \nonumber \\
&& \int dx_A dx_B \left\{ \sum_{ij}\left[ G_{i/A}(x_A,\mu_f)
G_{j/B}(x_B,\mu_f) \hat{\sigma}^{ij}_{NLO}(x_{A}x_{B}s,\mu_r)\right]
+(A\leftrightarrow B)\right\},
\end{eqnarray}
where the notations of $\mu_f$, $x_{A}$, $x_{B}$ are the same as
those in Eq.(\ref{sigma_PP}), but we adopt the CTEQ6m
PDFs\cite{cteq} for $G_{i/A}(x_{A},x_{B},\mu_f)$ and
$G_{j/B}(x_{A},x_{B},\mu_f)$ in the NLO calculations. The total NLO
QCD corrected cross section for the partonic process $kl \to W^- H^0
j$ can be expressed as
\begin{eqnarray}
\hat\sigma^{kl}_{NLO}=\hat\sigma^{kl}_{LO}
+\Delta\hat\sigma^{kl}_{NLO}=\hat\sigma^{kl}_{LO}+
\hat{\sigma}^{kl}_{R}+\hat{\sigma}^{kl}_{V},
\end{eqnarray}
For simplicity, we define the factorization and renormalization
scales being equal, i.e., $\mu_f=\mu_r=\mu$. At the Tevatron the
incoming colliding particles are proton and antiproton; the cross
sections for both the $p\bar p \to W^-H^0j+X$ and $p\bar p \to
W^+H^0j+X$ processes should be the same. In the following we
provide only the results for the former process. On the contrary,
we give the LHC results for the $pp \to W^-H^0j+X$ and $pp \to
W^+H^0j+X$ processes separately because of its proton-proton
colliding mode.

\par
\section{Numerical results and discussion}
\par
In our numerical calculations we take one-loop and two-loop
running $\alpha_{s}$ in the LO and NLO calculations,
respectively\cite{hepdata}. The QCD parameters are taken as
$\Lambda_5^{LO}=165~MeV$, $\Lambda_5^{\overline{MS}}=226~MeV$,
$N_f=5$. We take the renormalization and factorization scales to
be a common value as $\mu_r=\mu_f=\mu_0\equiv\frac{1}{2}(m_W+m_H)$
and $m_H=120~GeV$ by default. The colliding energies in the
proton-(anti)proton center-of-mass system are taken as $\sqrt
s=14~TeV$ for the LHC and $\sqrt s=1.96~TeV$ for the Tevatron Run
II. We set the values of the CKM matrix elements as
\begin{eqnarray}\label{CKM}
 V_{CKM} &=& \left(
\begin{array}{ccc}
    V_{ud} \ &  V_{us} \ &  V_{ub} \\
    V_{cd} \ &  V_{cs} \ &  V_{cb} \\
    V_{td} \ &  V_{ts} \ &  V_{tb} \\
\end{array}
    \right)=\left(
\begin{array}{ccc}
     0.97418 \ &  0.22577 \ &  0 \\
    -0.22577 \ &  0.97418 \ &  0 \\
       0 \ &  0 \ &  1 \\
\end{array}  \right).
\end{eqnarray}
The cosine of the weak mixing angle squared is set to its on-shell
value obtained by $c_{W}^{2}=m_{W}^{2}/m_{Z}^{2}$. The weak vector
boson and top-quark masses are taken as $m_W = 80.398~GeV$, $m_Z =
91.1876~GeV$ and $m_t=171.2~GeV$. The fine structure constant at
the $Z^0$-pole has the value as
$\alpha(m_Z^2)^{-1}=127.925$\cite{hepdata}.

\par
In the LO and NLO calculations we adopt the massless five-flavor
scheme and put the restriction of $p_{T}^{j}>p_{T,j}^{cut}$ on the
jet transverse momentum for one-jet events. For the two-jet events
(originating from the real corrections), we apply the jet
algorithm of Ref.\cite{jet} in the definition of the tagged hard
jet with $R=1$. That means when two jets in the final state
satisfy the constraint of $\sqrt{\Delta \eta^2 + \Delta \phi^2} <
R \equiv 1$(where $\Delta \eta$ and $\Delta \phi$ are the
differences of rapidity and azimuthal angle between the two jets),
we merge them into one new ``jet'' and consider it as an one-jet
event. In handling the one- and two-jet events we use the so
called ``inclusive'' scheme in default of other statement. In this
scheme we demand $p_{T}^{j}> p_{T,j}^{cut}$ for the one-jet
events, and for the two-jet events we apply the constraint of
$p_{T}^{j}> p_{T,j}^{cut}$ on the leading jet but not on the
second jet, where the leading jet and the second jet are
characterized by $E_{T}({the~leading~jet})
> E_{T}({the~second~jet})$.

\par
Since the events involving the final hard $b(\bar{b})$-jet can be
experimentally excluded by anti b-tagging, we consider only the
phase-space with $b\bar b$ jets satisfying $\sqrt{\Delta
\eta^2+\Delta \phi^2}<1$ for each partonic $W^{\pm}H^0b\bar b$
production process. For these events, the final $b$ and $\bar b$
are accepted as one hard ``jet'' when its transverse momentum
$p_{T}^{j}>p_{T,j}^{cut}$.

\par
\begin{figure}[htbp]
\includegraphics[scale=0.8]{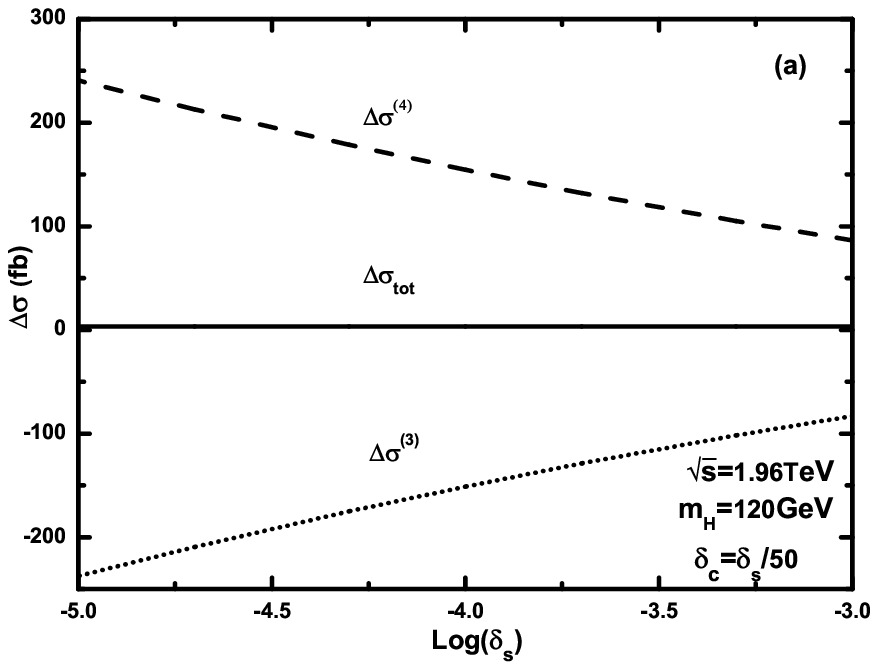}
\includegraphics[scale=0.8]{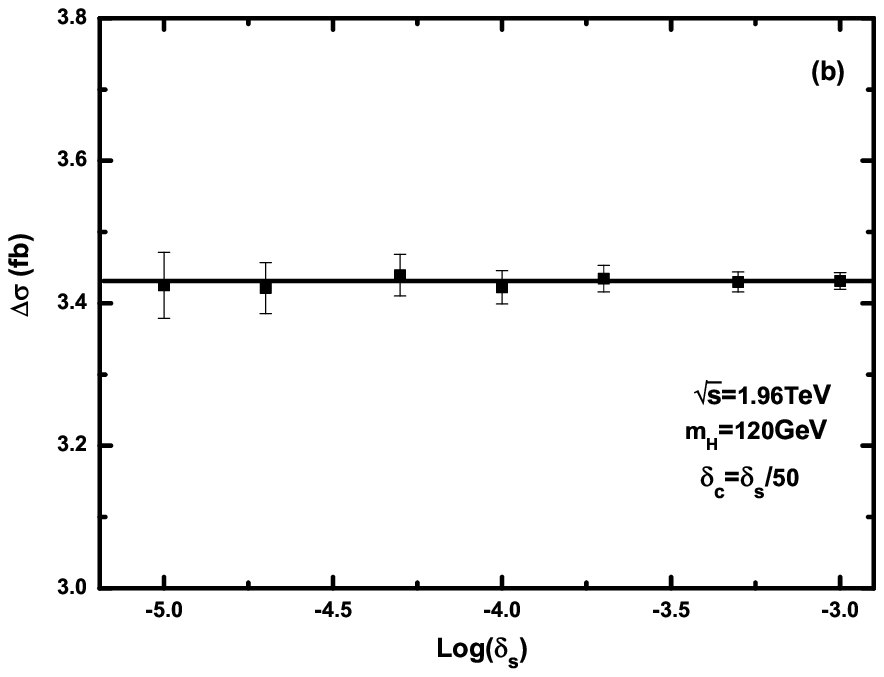}
\hspace{0in}%
\caption{\label{fig3} (a) the dependence of the NLO QCD
corrections to the $p\bar p \to W^-H^0j+X$ process on the soft
cutoff $\delta_s$ and $\delta_c$ at the Tevatron, where we take
$\delta_c=\delta_s/50$, $\mu=\mu_0$, $m_H=120~GeV$, and apply the
inclusive scheme with $p_{T,j}^{cut}=20~GeV$. (b) the amplified
curve for $\Delta\sigma_{NLO}$ of Fig.\ref{fig3}(a). }
\end{figure}

In order to verify the correctness of our results, we made
following verifications:

\begin{enumerate}
\item
The UV and IR safeties are verified numerically after combining all
the contributions at the NLO.

\item
The LO cross section for the process $p\bar p \to \bar u d \to
W^-H^0j+X$ at the Tevatron was calculated by using two independent
developed programs: FeynArts3.4/FormCalc5.4 \cite{fey,formloop} and
CompHEP-4.4p3 programs\cite{CompHEP}, and applying the Feynman and
unitary gauges separately. The results are in agreement within the
statistic errors. The virtual correction and the real
gluon/light-(anti)quark correction to the $pp \to W^-H^0j+X$ process
at the LHC were evaluated twice independently based on different
codes, and yield results in mutual agreement.

\item
The total NLO QCD correction being independent of the two cutoffs,
$\delta_s$ and $\delta_c$, has been numerically verified. In
Figs.\ref{fig3}(a) and (b) we depict the total NLO QCD corrections
to the $p\bar p \to W^-H^0j+X$ process at the Tevatron as the
functions of the cutoffs $\delta_s$ and $\delta_c$. There we apply
the inclusive scheme with $p_{T,j}^{cut}=20~GeV$, and take
$\mu=\mu_0$, $m_H=120~GeV$, $\delta_c = \delta_s/50$. The
amplified curve for $\Delta\sigma_{NLO}$ of Fig.\ref{fig3}(a) is
presented in Fig.\ref{fig3}(b) together with calculation errors.
The figures demonstrate that the total NLO QCD correction does not
depend on the arbitrarily chosen value of the cutoff
$\delta_s$($\delta_c$) within statistic errors. Figure \ref{fig3}(a)
shows that although the three-body
correction($\Delta\sigma^{(3)}$) and four-body
correction($\Delta\sigma^{(4)}$) are strongly related with the
cutoff $\delta_s$($\delta_c$), the final total NLO QCD correction
$\Delta\sigma_{NLO}$ which is the summation of the three-body term
and four-body term, i.e.,
$\Delta\sigma_{NLO}=\Delta\sigma^{(3)}+\Delta\sigma^{(4)}$, is
independent of the two cutoffs within the statistic errors. The
independence of the full NLO QCD corrections to the $p \bar{p} \to
W^- H^0 j + X$ process on the cutoffs $\delta_s$ and $\delta_c$
provides an indirect check for the correctness of the
calculations. In further numerical calculations, we fix
$\delta_s=5 \times 10^{-4}$ and $\delta_c=\delta_s/50$.
\end{enumerate}

\par
\begin{figure}[htbp]
\includegraphics[scale=0.75]{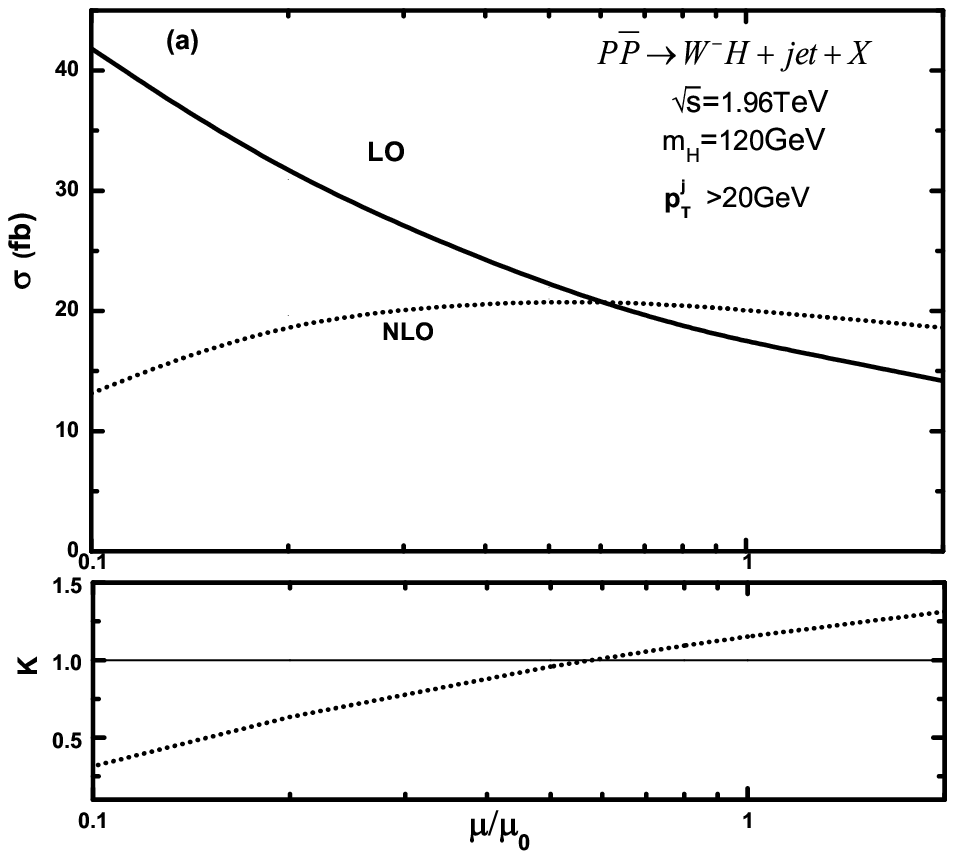}
\includegraphics[scale=0.75]{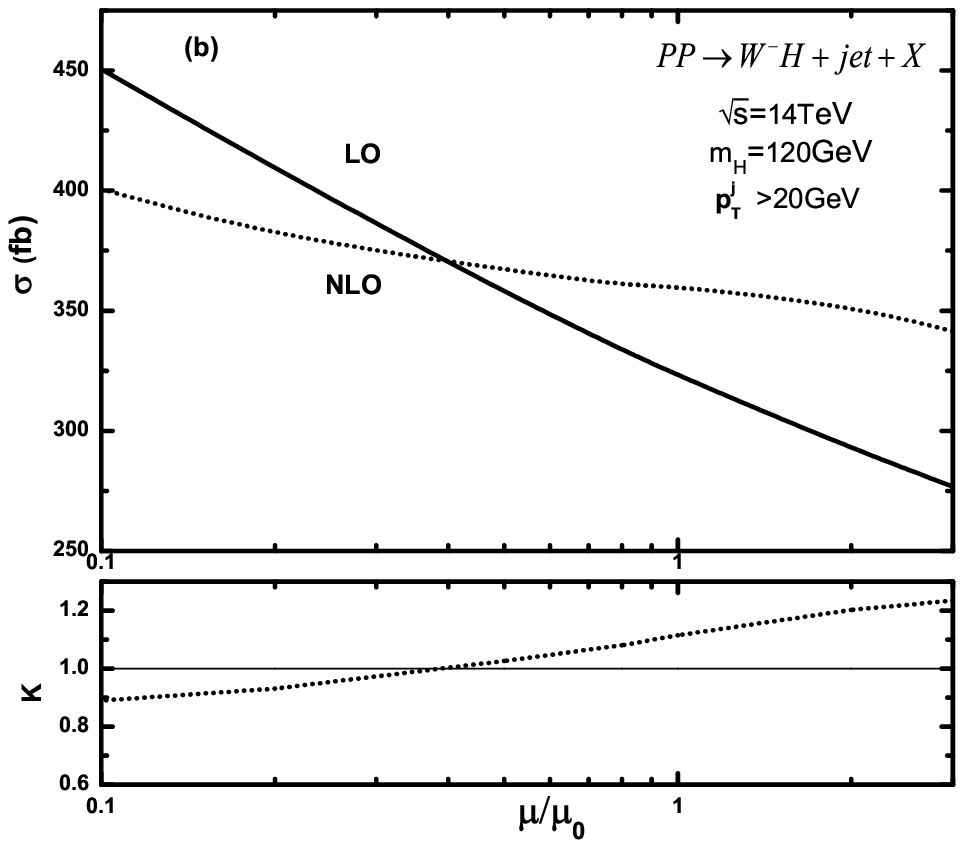}
\includegraphics[scale=0.75]{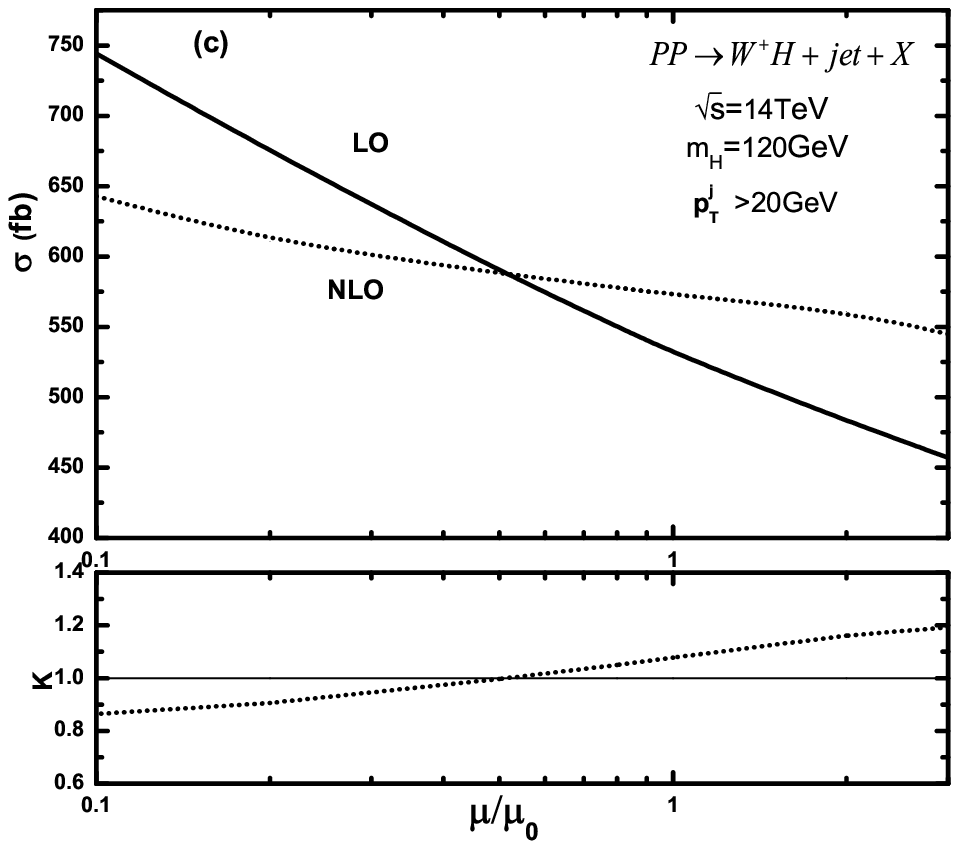}
\caption{\label{fig4} The LO, NLO corrected cross sections and the
corresponding K-factor
($K(\mu)\equiv\sigma_{NLO}(\mu)/\sigma_{LO}(\mu)$) versus the
factorization/renormalization scale($\mu=\mu_r=\mu_f$) by applying
the inclusive scheme with $p_{T,j}^{cut}=20~GeV$. (a) for \ppbwhj
at the Tevatron, (b) for \ppwhjm at the LHC, (c) For \ppwhjp at
the LHC. }
\end{figure}

In Figs.\ref{fig4}(a), (b) and (c) we present the scale dependence
of the LO, NLO cross sections, and the corresponding
K-factor($K(\mu)\equiv \sigma_{NLO}(\mu)/\sigma_{LO}(\mu)$) upon
varying renormalization and factorization scales in the
$\mu\equiv\mu_r=\mu_f$ way for the $p\bar p \to W^-H^0j+X$ process
at the Tevatron and the $pp \to W^{\mp}H^0j+X$ processes at the
LHC, separately. There we adopt the inclusive scheme with
$p_{T,j}^{cut}=20~GeV$. Figure \ref{fig4}(a) shows that when the
curve transits from LO to the NLO, the scale uncertainty (defined
in the range of $0.5\mu_0 < \mu <2\mu_0$) is reduced by the NLO
QCD corrections from $44.6\%$(LO) to $11.9\%$(NLO) at the
Tevatron. Figures \ref{fig4}(b,c) show that the scale
uncertainties(defined in the range of $0.5\mu_0 < \mu <2\mu_0$) at
the LHC are reduced from $20.1\%$(LO) to $4.2\%$(NLO) for the
\ppwhjm process and from $20.0\%$(LO) to $4.7\%$(NLO) for the
\ppwhjp process, respectively. We can read from these figures that
the K-factor($K(\mu)\equiv \sigma_{NLO}(\mu)/\sigma_{LO}(\mu)$)
for the process $p\bar p \to W^-H^0j+X$ at the Tevatron is in the
range of $[0.31, 1.31]$, while the K-factors for the processes $pp
\to W^-H^0j+X$ and $pp \to W^+H^0j+X$ at the LHC, vary in the
ranges of $[0.89, 1.23]$ and $[0.86, 1.19]$ in the plotted
$\mu/\mu_0$ ranges, respectively.

\par
\begin{figure}[htbp]
\includegraphics[scale=0.75]{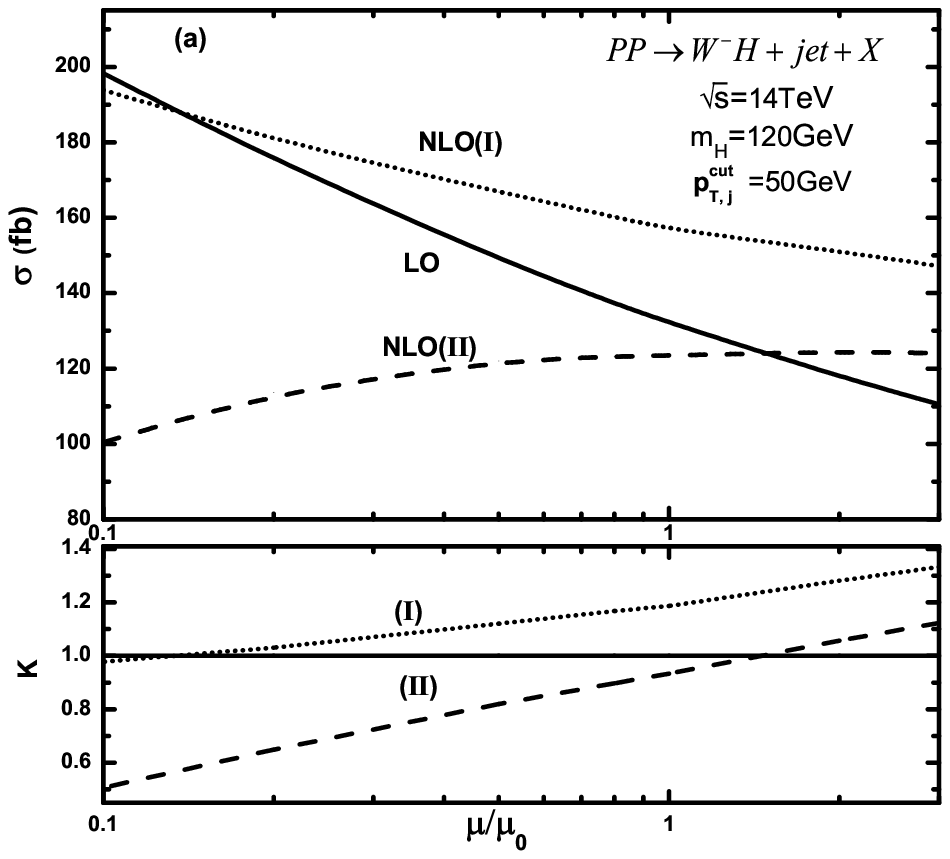}
\includegraphics[scale=0.75]{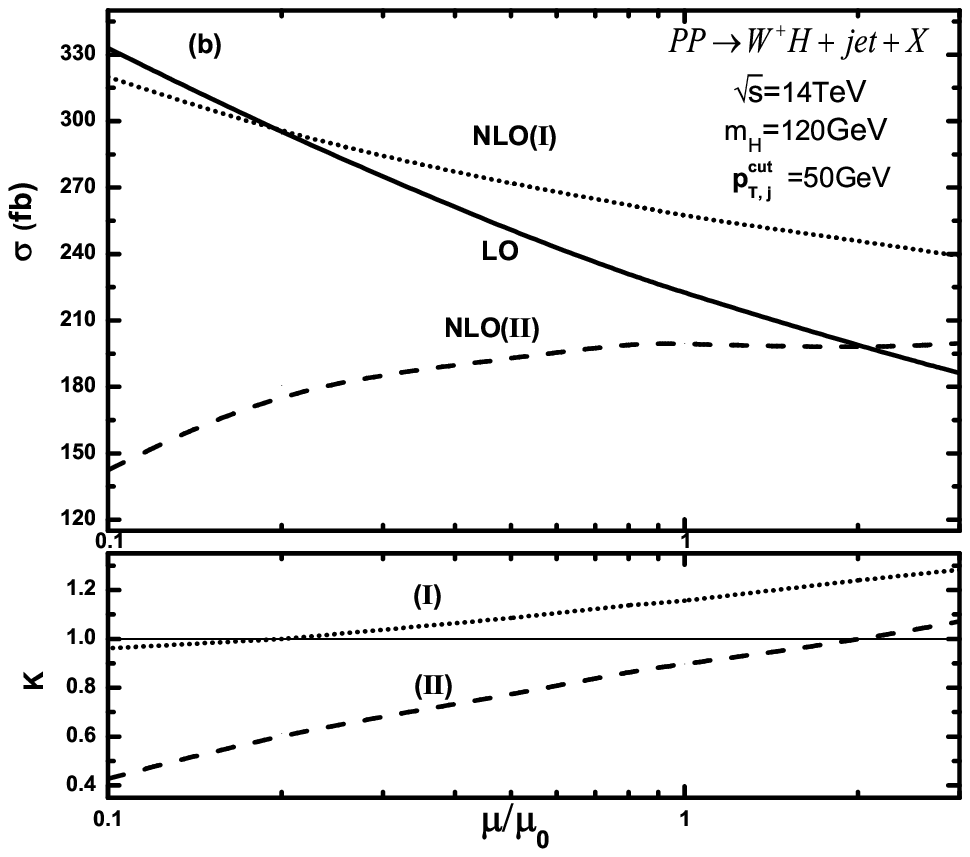}
\caption{\label{fig5} The LO, NLO corrected cross sections and the
corresponding K-factor($K(\mu)\equiv
\sigma_{NLO}(\mu)/\sigma_{LO}(\mu)$) at the LHC by taking
$p_{T,j}^{cut}=50~GeV$ and adopting separately (I) the inclusive
scheme and (II) exclusive scheme. (a) for \ppwhjm, (b) for
\ppwhjp.  }
\end{figure}

In our NLO calculation, we find that the LO, NLO QCD corrected
cross sections and the QCD K-factors for the \ppwhjm and \ppwhjp
processes at the LHC are sensitive to both the transverse momentum
cut on the leading jet, $p_{T,j}^{cut}$, and the jet event
selection scheme. In order to demonstrate this influence, we
present the LO, NLO QCD corrected cross sections and the QCD
K-factors for \ppwhjm and \ppwhjp processes at the LHC with
$p_{T,j}^{cut} = 50~GeV$ in Figs.5(a) and (b), respectively. The
curves labeled with ``NLO(I)'' and ``NLO(II)'' correspond to the NLO
QCD corrected cross sections with two different jet event
selection schemes: (i) the ``inclusive'' scheme as declared above;
(ii) the ``exclusive'' scheme, which means the one-jet events with
$p_{T}^{j}>p_{T,j}^{cut}$ are accepted, and the two-jet events
with $p_{T}^{j}$(the second hard jet)$> p_{T,j}^{cut}$ are
vetoed\cite{Uwer}. We can see from Figs.5(a) and (b) that the NLO
QCD corrections can significantly reduce the
factorization/renormalization scale dependence. By adopting the
inclusive selection scheme, the LHC LO scale uncertainty(defined
in the range of $0.5\mu_0 < \mu <2\mu_0$) for the \ppwhjm process
(the \ppwhjp process) is about $23.5\%$($23.3\%$), and is reduced
to about $10.1\%$($9.91\%$) by the NLO QCD corrections.
Alternatively when a veto against the emission of a second hard
jet is applied (i.e., by adopting the exclusive scheme), the LHC
scale uncertainty(defined in the range $0.5\mu_0 < \mu <2\mu_0$)
of the \ppwhjm process(the \ppwhjp process) is improved by the NLO
QCD correction to the value of $2.06\%$($3.53\%$). It shows that
the reduction of the scale uncertainty by the exclusive LHC NLO
correction is larger than the inclusive NLO correction. From
Figs.\ref{fig5}(a,b) we can see that by taking
$p_{T,j}^{cut}=50~GeV$, the exclusive LHC NLO cross section for
the \ppwhjm process(or the \ppwhjp process) decreases in the low
scale region as shown by the curves labeled with ``NLO(II)''.
Therefore, we can conclude that the curve feature of the LHC NLO
QCD corrected cross section for the \ppwhjm or the \ppwhjp process
versus scale $\mu$ is correlated to the $p_{T,j}^{cut}$ value and
the jet event selection scheme.

\par
In Table \ref{tab1} we list some of the representative numerical
results for the LO, NLO corrected cross sections and their
corresponding K-factors by applying the inclusive scheme with
$p_{T,j}^{cut}=20~GeV$, and taking $m_H=120~GeV$, the energy scale
$\mu=0.5\mu_0$, $\mu_0$, $2\mu_0$, $\mu_1$ and $\mu_2$ separately,
where $\mu_1$ and $\mu_2$ are phase-space dependent scales defined
as $\mu_1\equiv
\sqrt{\frac{1}{2}\left(\left(p_T^{W}\right)^2+\left(p_T^{H}\right)^2+
m_W^2+m_H^2\right)}$ and $\mu_2\equiv
\sqrt{\left(p_T^{W}\right)^2+\left(p_T^{H}\right)^2+
m_W^2+m_H^2}$.
\begin{table}
\begin{center}
\begin{tabular}{|c|c|c|c|c|}
\hline  Process          &$\mu (GeV)$& $\sigma_{LO}(fb)$ & $\sigma_{NLO}(fb)$ & $K$ \\
\hline                               &$0.5\mu_0$ & 21.949(3)  & 21.01(2)   & 0.96   \\
\cline{2-5}$p\bar{p}\to W^{-}H^0j+X$ &$\mu_0$    & 17.440(2)  & 20.08(2)   & 1.15   \\
\cline{2-5}                          &$2\mu_0$   & 14.167(2)  & 18.61(1)   & 1.31   \\
\cline{2-5}$\sqrt{s}=1.96~TeV$       &$\mu_1$    & 16.0128(8) & 19.60(1)   & 1.22   \\
\cline{2-5}                          &$\mu_2$    & 14.457(1)  & 18.79(1)   & 1.30   \\
\hline                               &$0.5\mu_0$ & 357.58(2)  & 367.2(2)   & 1.03   \\
\cline{2-5}$pp\to W^{-}H^0j+X$       &$\mu_0$    & 323.03(2)  & 360.2(2)   & 1.12   \\
\cline{2-5}                          &$2\mu_0$   & 292.76(2)  & 352.1(2)   & 1.20   \\
\cline{2-5}$\sqrt{s}=14~TeV$         &$\mu_1$    & 306.023(8) & 350.9(1)   & 1.15   \\
\cline{2-5}                          &$\mu_2$    & 291.63(1)  & 347.9(1)   & 1.19   \\
\hline                               &$0.5\mu_0$ & 589.49(5)  & 588.0(3)   & 0.997   \\
\cline{2-5}$pp\to W^{+}H^0j+X$       &$\mu_0$    & 531.37(3)  & 572.9(3)   & 1.08   \\
\cline{2-5}                          &$2\mu_0$   & 483.21(4)  & 561.0(3)   & 1.16   \\
\cline{2-5}$\sqrt{s}=14~TeV$         &$\mu_1$    & 503.36(2)  & 561.2(2)   & 1.12   \\
\cline{2-5}                          &$\mu_2$    & 479.93(2)  & 556.4(2)   & 1.16   \\
\hline
\end{tabular}
\end{center}
\begin{center}
\begin{minipage}{15cm}
\caption{\label{tab1} The numerical results for the LO, NLO QCD
corrected cross sections and their corresponding
K-factors($K(\mu)\equiv \sigma_{NLO}(\mu)/\sigma_{LO}(\mu)$) by
applying the inclusive scheme with $p_{T,j}^{cut}=20~GeV$, and
taking $m_H=120~GeV$ and different values of scale $\mu$ for the
process $p\bar p \to W^-H^0j+X$ at the Tevatron Run II, the
processes $pp \to W^-H^0j+X$ and $pp \to W^+H^0j+X$ at the LHC. In
this table we denote $\mu_0=\frac{1}{2}(m_W+m_H)$,
$\mu_1=\sqrt{\frac{1}{2}\left[
\left(p_T^{W}\right)^2+\left(p_T^{H}\right)^2+m_W^2+m_H^2\right]}$
and $\mu_2=\sqrt{\left(p_T^{W}\right)^2+
\left(p_T^{H}\right)^2+m_W^2+m_H^2}$. }
\end{minipage}
\end{center}
\end{table}
\par
In Table \ref{tab2}, we list some of the numerical results of the
LO and the QCD corrected cross sections and the corresponding
K-factors($K\equiv\frac{\sigma_{NLO}}{\sigma_{LO}}$) for the
$p\bar{p} \to W^{-}H^0j+X$ process at the Tevatron and the $pp \to
W^{\pm}H^0j+X$ processes at the LHC, where we apply the inclusive
scheme with $p_{T,j}^{cut}=20~GeV$, and take $\mu=\mu_0$, the
values of Higgs-boson mass as $120~GeV$, $150~GeV$ and $180~GeV$,
separately. Table \ref{tab2} shows both the LO and NLO QCD
corrected cross sections and K-factors are all sensitive to the
Higgs mass. Among them the K-factor for the $pp \to W^{+}H^0j+X$
process at the LHC is less sensitive to the Higgs boson mass than
others. We also find the LO and the NLO QCD corrected cross
sections decrease rapidly with the increment of $m_{H}$ at both
hadronic colliders.
\begin{table}
\begin{center}
\begin{tabular}{|c|c|c|c|c|}
\hline  process &$m_H (GeV)$ & $\sigma_{LO}(fb)$ & $\sigma_{NLO}(fb)$ & $K$  \\
\hline   $p\bar{p}\to W^{-}H^0j+X$    &120  &17.440(2)  & 20.08(2)  & 1.15   \\
\cline{2-5}                           &150  & 8.2697(8) & 9.306(7)  & 1.13   \\
\cline{2-5} $\sqrt{s}=1.96TeV$        &180  & 4.2729(4) & 4.629(3)  & 1.08   \\
\hline  $pp\to W^{-}H^0j+X$           &120  & 323.03(2) & 360.2(2)  & 1.12   \\
\cline{2-5}                           &150  & 164.96(1) & 180.94(8) & 1.10   \\
\cline{2-5} $\sqrt{s}=14TeV$          &180  & 93.692(6) & 100.26(4) & 1.07   \\
\hline  $pp\to W^{+}H^0j+X$           &120  & 531.37(3) & 572.9(3)  & 1.08   \\
\cline{2-5}                           &150  & 284.49(2) & 294.2(2)  & 1.03   \\
\cline{2-5} $\sqrt{s}=14TeV$          &180  & 166.18(1) & 167.9(1)  & 1.01   \\
\hline
\end{tabular}
\end{center}
\begin{center}
\begin{minipage}{14cm}
\caption{\label{tab2} The numerical results for the LO and the NLO
QCD corrected cross sections and the corresponding K-factor
($K\equiv\frac{\sigma_{NLO}}{\sigma_{LO}}$) with $\mu=\mu_0$,
$m_{H}=120~GeV$, $150~GeV$ and $180~GeV$, for the $p\bar{p} \to
W^{-}H^0j+X$ process at the Tevatron and the $pp \to
W^{\pm}H^0j+X$ processes at the LHC by applying the inclusive
scheme with $p_{T,j}^{cut}=20~GeV$. }
\end{minipage}
\end{center}
\end{table}

\par
\begin{figure}
\centering
\includegraphics[scale=0.75]{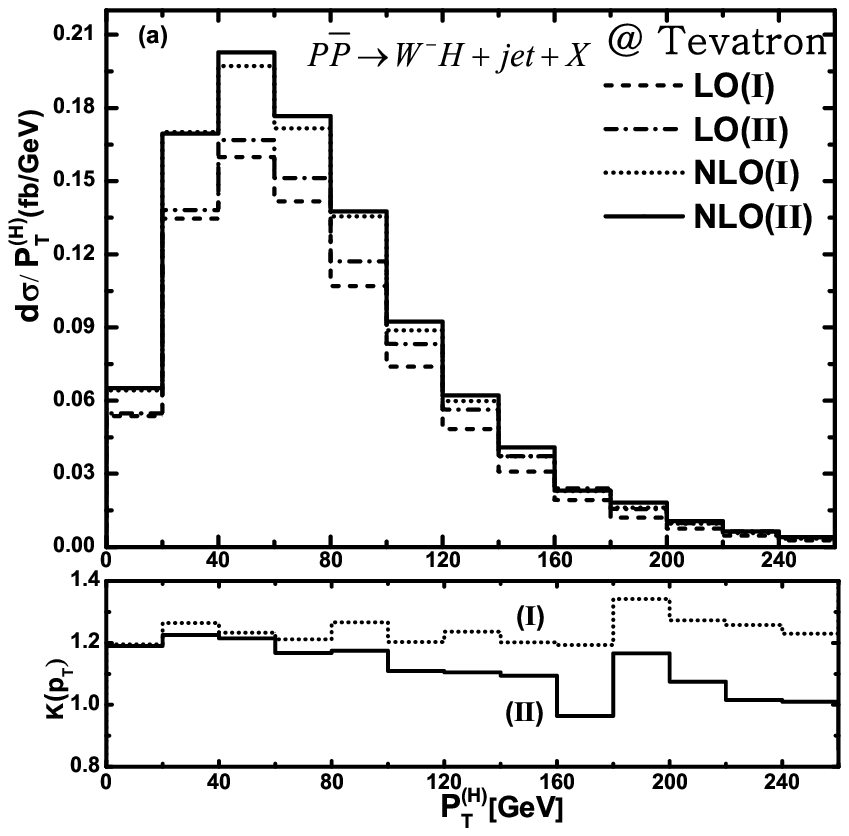}
\includegraphics[scale=0.75]{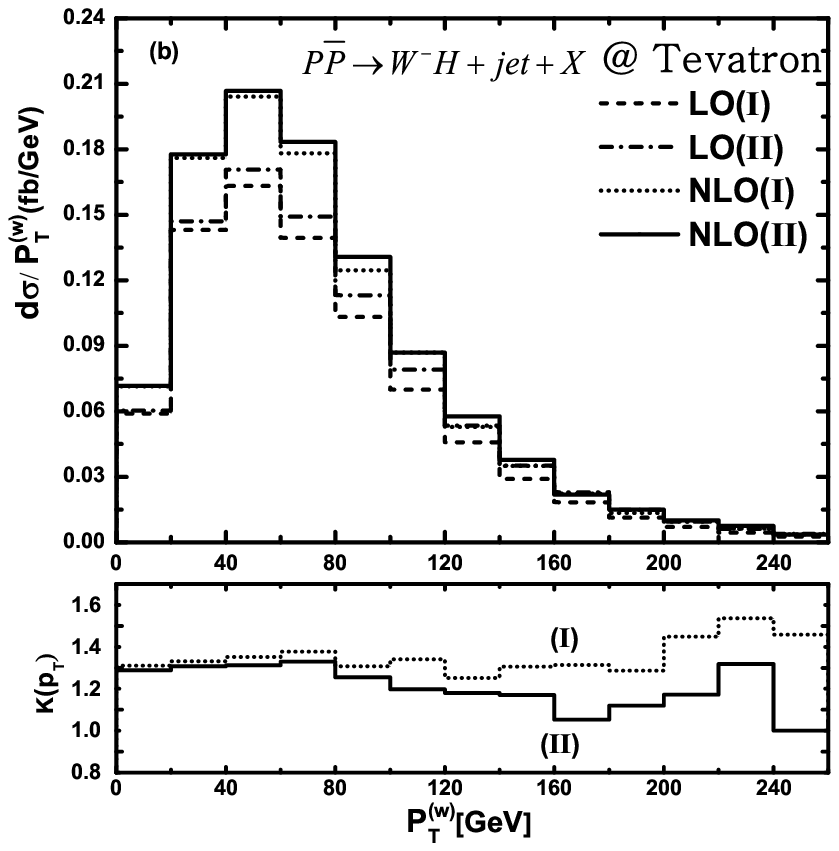}
\includegraphics[scale=0.75]{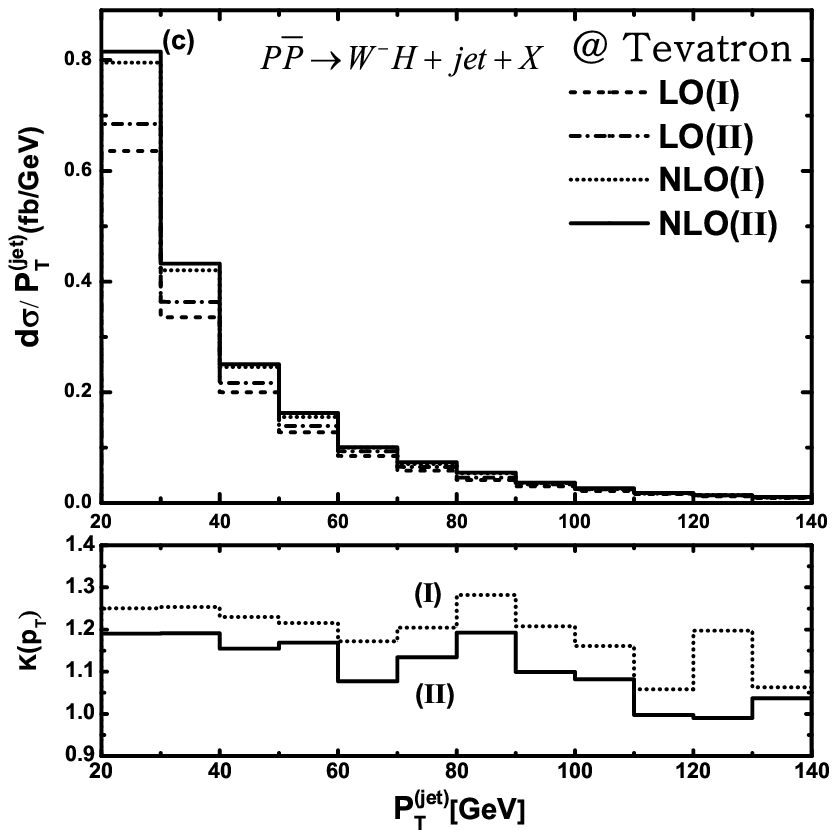}
\caption{\label{fig6} The LO and NLO QCD corrected distributions
of the transverse momenta of final particles and corresponding
K-factors ($K(p_T)\equiv \frac{d \sigma_{NLO}}{dp_T}/ \frac{d
\sigma_{LO}}{dp_T}$) for the process $p\bar{p} \to W^-H^0j+X$ at
the Tevatron with $m_H=120~GeV$. The distributions labeled by (I)
and (II) are for the $\mu=\mu_1$ and $\mu=\mu_0$ respectively. (a)
for $H^0$-boson, (b) for $W^{-}$-boson, (c) for final leading jet.
}
\end{figure}
\begin{figure}
\centering
\includegraphics[scale=0.75]{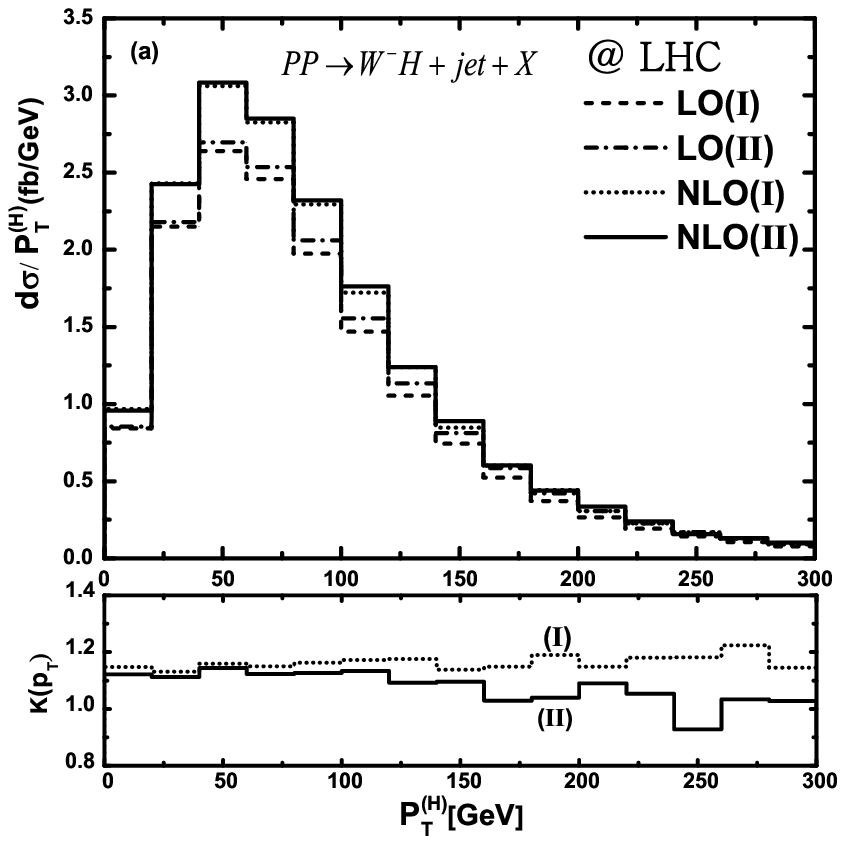}
\includegraphics[scale=0.75]{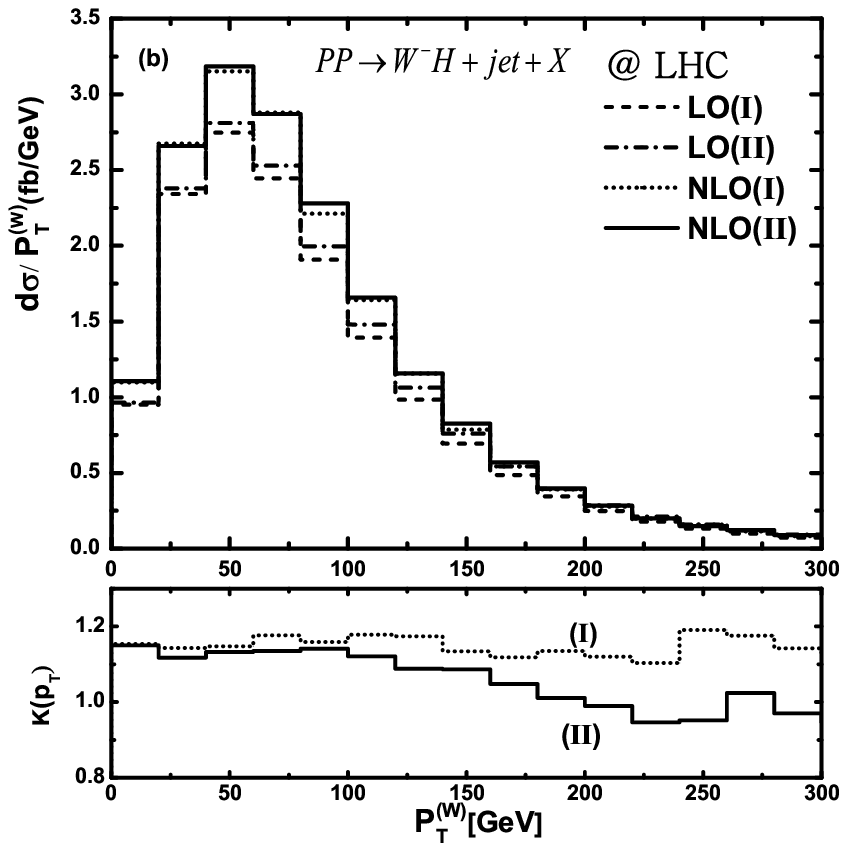}
\includegraphics[scale=0.75]{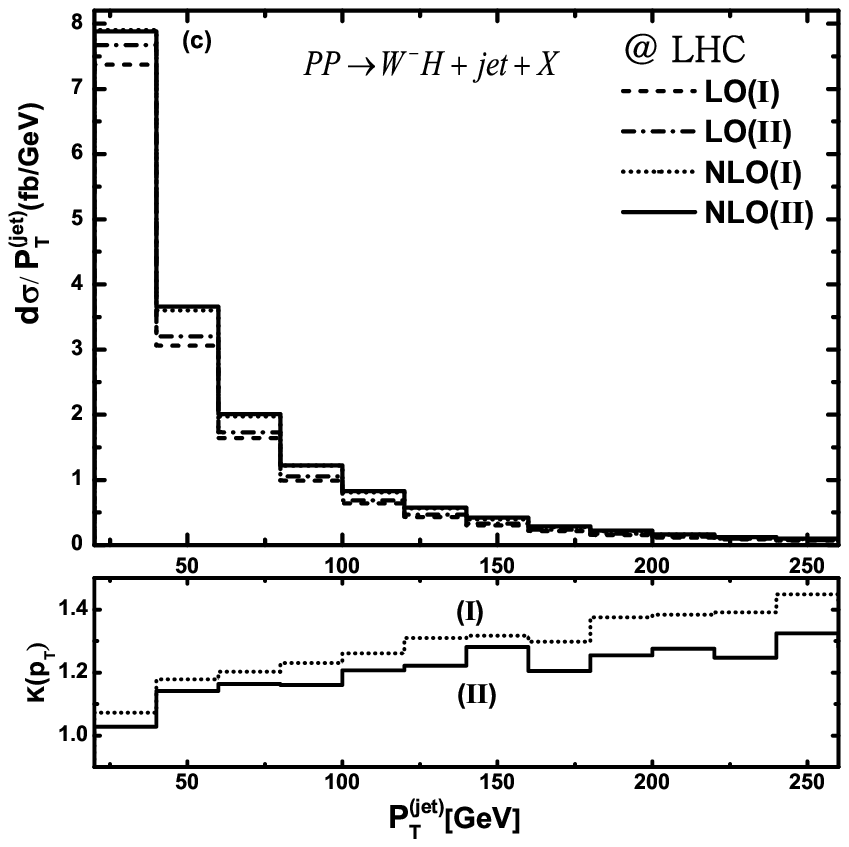}
\caption{\label{fig7} The LO and NLO QCD corrected distributions
of the transverse momenta of final particles and corresponding
K-factors ($K(p_T)\equiv \frac{d \sigma_{NLO}}{dp_T}/ \frac{d
\sigma_{LO}}{dp_T}$) for the process \ppwhjm at the LHC with
$m_H=120~GeV$. The distributions labeled by (I) and (II) are for
the $\mu=\mu_1$ and $\mu=\mu_0$ respectively. (a) for $H^0$-boson,
(b) for $W^{-}$-boson, (c) for final leading jet.   }
\end{figure}
\begin{figure}
\centering
\includegraphics[scale=0.75]{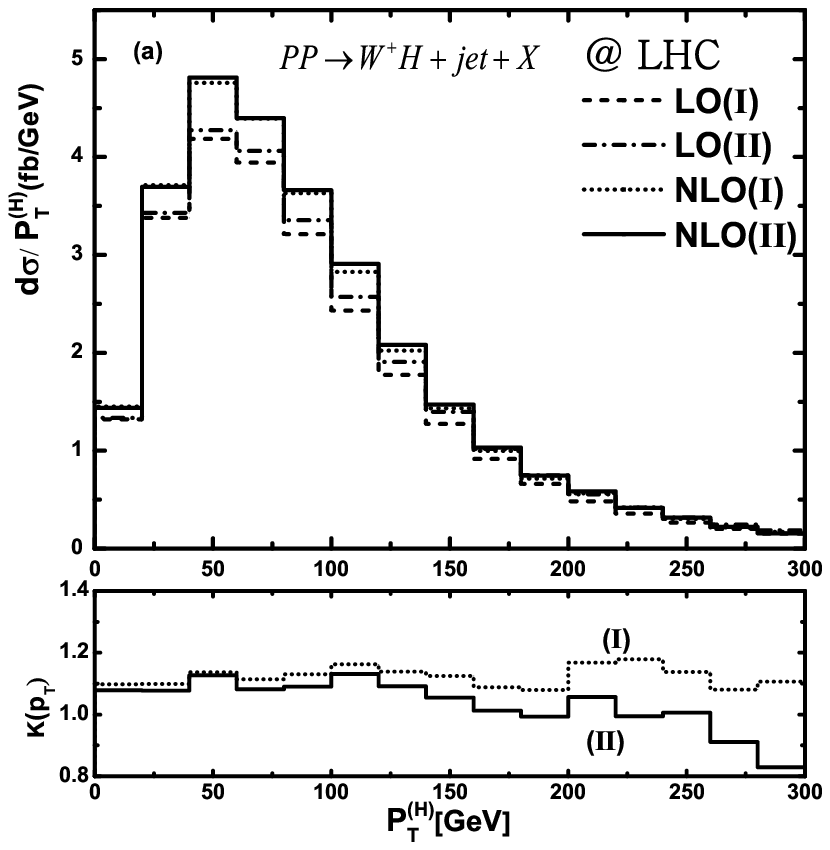}
\includegraphics[scale=0.75]{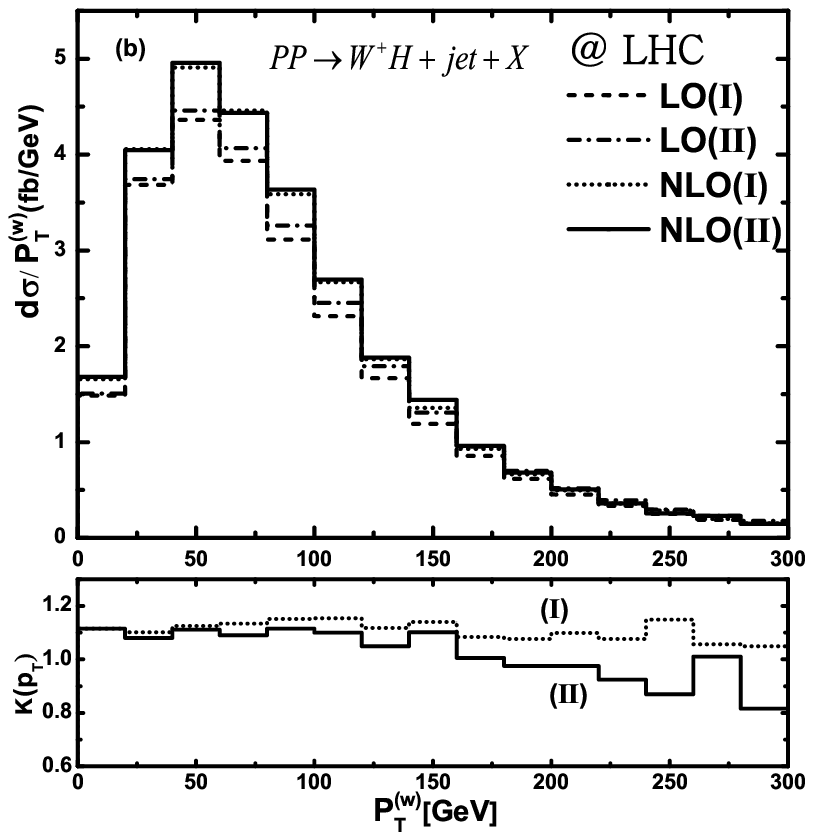}
\includegraphics[scale=0.75]{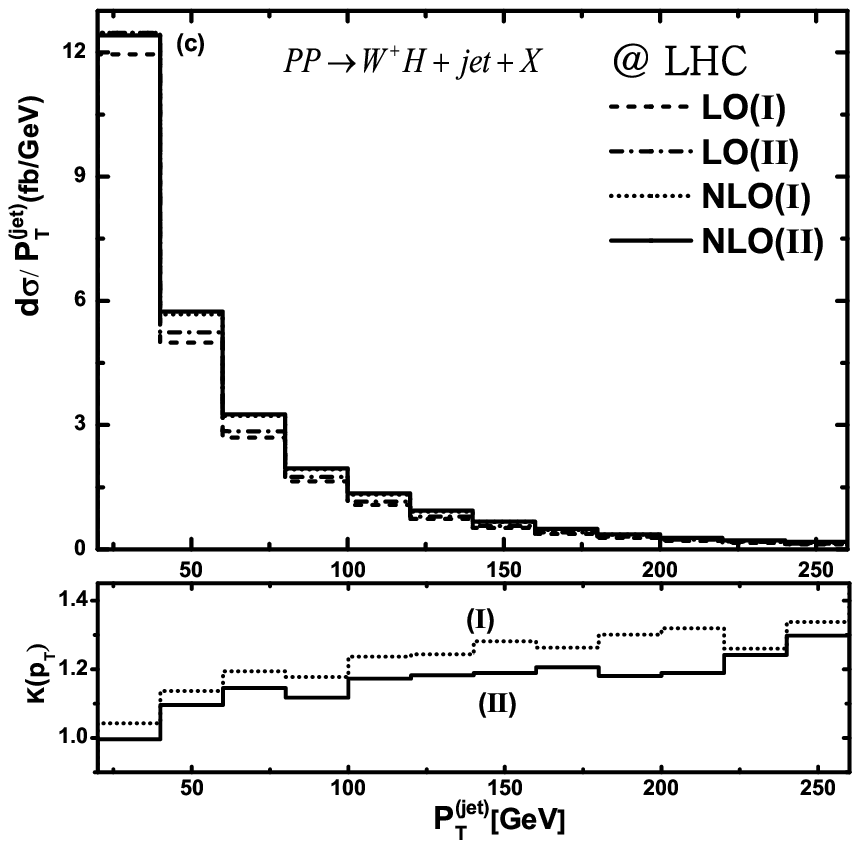}
\caption{\label{fig8} The LO and NLO QCD corrected distributions
of the transverse momenta of final particles and corresponding
K-factors ($K(p_T)\equiv \frac{d \sigma_{NLO}}{dp_T}/ \frac{d
\sigma_{LO}}{dp_T}$) for the process \ppwhjp at the LHC with
$m_H=120~GeV$. The distributions labeled by (I) and (II) are for
the $\mu=\mu_1$ and $\mu=\mu_0$ respectively. (a) for $H^0$-boson,
(b) for $W^{+}$-boson, (c) for final leading jet.  }
\end{figure}

In Figs.\ref{fig6}(a,b,c) we depict the LO and NLO QCD corrected
differential cross sections of the transverse momenta for the
final produced $H^0$-, $W^-$-boson and leading jet in the process
$p\bar p \to W^-H^0j+X$ at the Tevatron, and the corresponding
K-factors($K(p_T)\equiv
\frac{d\sigma_{NLO}}{dp_T}/\frac{d\sigma_{LO}}{dp_T}$),
separately. There we adopt the inclusive scheme with
$p_{T,j}^{cut}=20~GeV$ and take $m_H=120~GeV$. The distributions
in these figures marked with (I) and (II) are for the $\mu=\mu_1$
and $\mu=\mu_0$ respectively. The analogous plots for the
processes \ppwhjm and \ppwhjp at the LHC are depicted in
Figs.\ref{fig7}(a,b,c) and Figs.\ref{fig8}(a,b,c), respectively.
In these figures we provide the NLO QCD corrected differential
cross sections($\frac{d\sigma_{NLO}}{dp_{T}}$) by applying the
inclusive scheme with $p_{T,j}^{cut}=20~GeV$. The plot for
$\frac{d\sigma_{NLO}}{dp_{T}^j}$ refers to the distribution of the
transverse momentum of the leading jet. Figures.\ref{fig6}(a,b),
Figs.\ref{fig7}(a,b) and Figs.\ref{fig8}(a,b) show that at both
the Tevatron and the LHC the NLO QCD corrections significantly
enhance the LO differential cross sections of $p_T^{H}$ and
$p_T^{W}$, especially when $p_T^{H}$, $p_T^{W}<150~GeV$. We
observe also that the curves for the $K(p_T^H)$- and
$K(p_T^W)$-factors in these figures become more stable in the
transition from $\mu=\mu_0$ to the phase-space dependent scale
$\mu=\mu_1$ (i.e., $\mu_1=\sqrt{\frac{1}{2}\left[
\left(p_T^{W}\right)^2+\left(p_T^{H}\right)^2+m_W^2+m_H^2\right]}$).
We can see from Fig.\ref{fig6}(c), Fig.\ref{fig7}(c) and
Fig.\ref{fig8}(c) that most of the leading jets are produced in the
low transverse momentum range, and the differential cross section
of $p_T^{j}$ is significantly enhanced by the NLO QCD corrections.

\par
\begin{figure}
\centering
\includegraphics[scale=0.65]{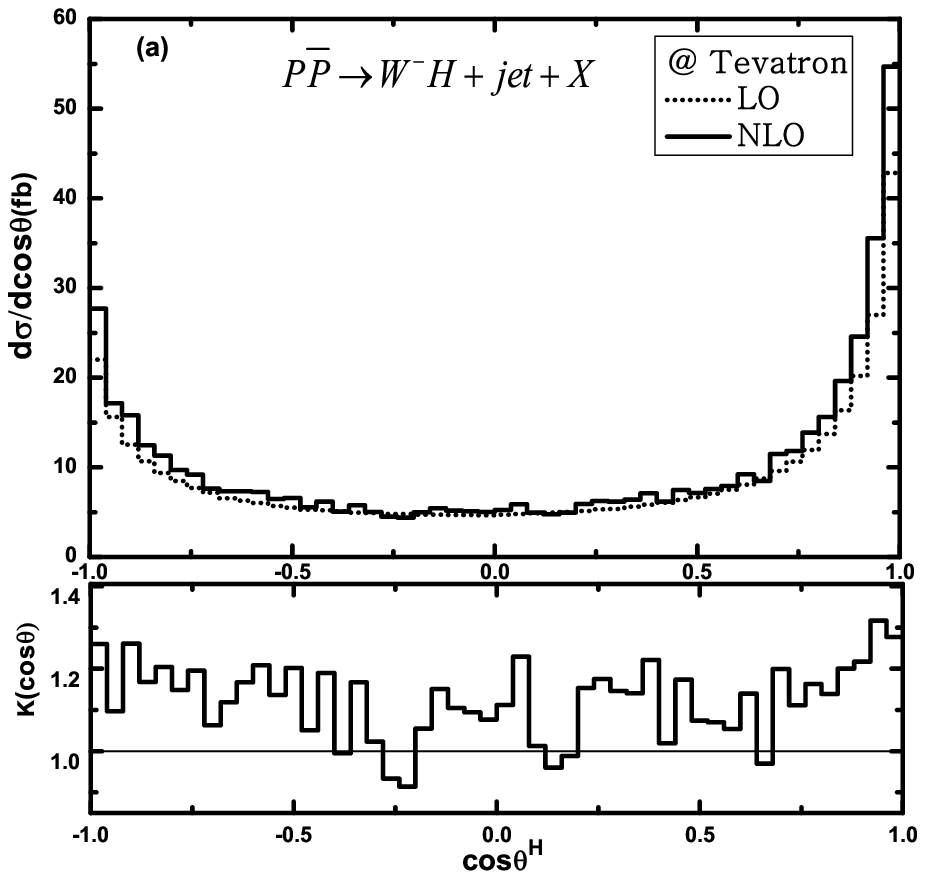}
\includegraphics[scale=0.65]{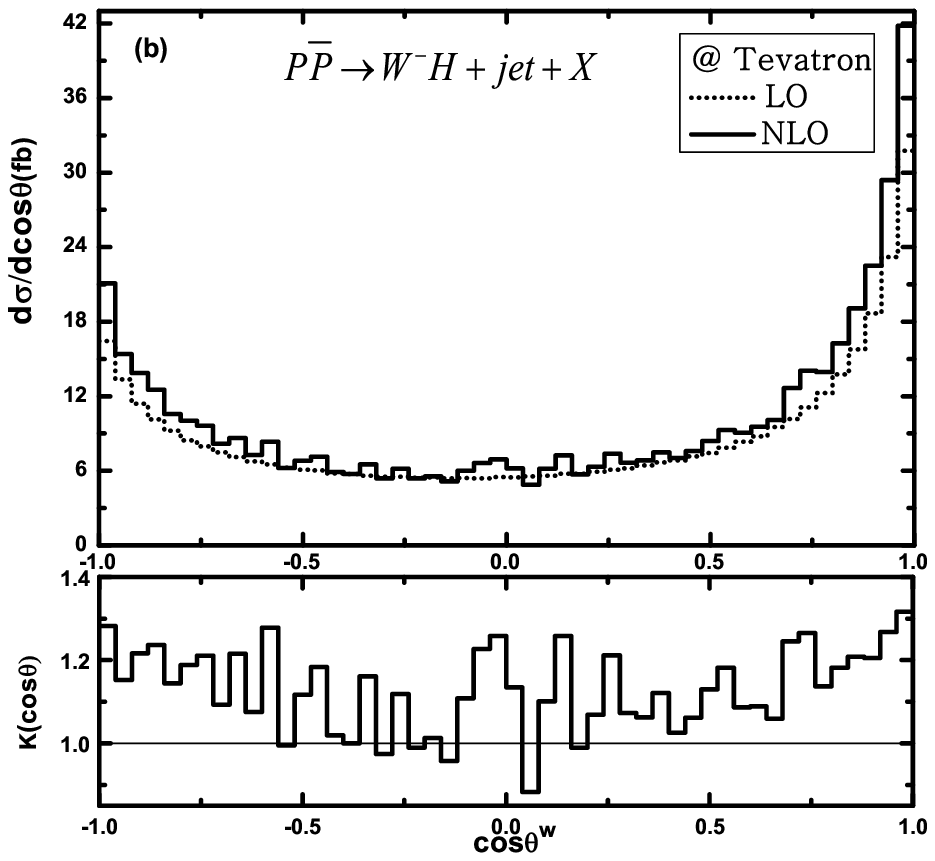}
\includegraphics[scale=0.65]{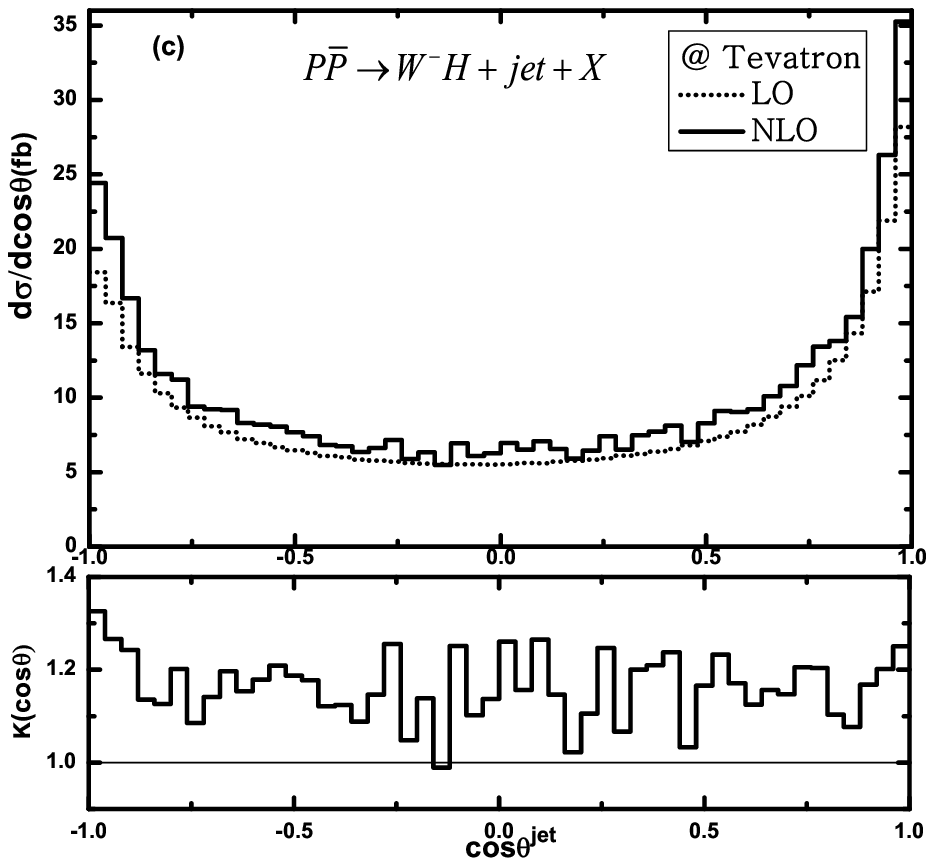}
\caption{\label{fig9} The LO, NLO QCD corrected
distributions($\frac{d\sigma}{d\cos\theta}$) and their corresponding
K-factors($K(\cos\theta)\equiv\frac{d\sigma_{NLO}}
{d\cos\theta}/\frac{d\sigma_{LO}} {d\cos\theta}$) versus the cosine
of the angle between the final particle and the direction of the
incoming antiproton for the process $p\bar p \to W^-H^0j+X$ at the
Tevatron by applying the inclusive scheme with
$p_{T,j}^{cut}=20~GeV$ and taking $\mu=\mu_0$ and $m_H=120~GeV$. (a)
for final Higgs-boson, (b) for final $W$-boson, (c) For final
leading jet. }
\end{figure}
\begin{figure}
\centering
\includegraphics[scale=0.65]{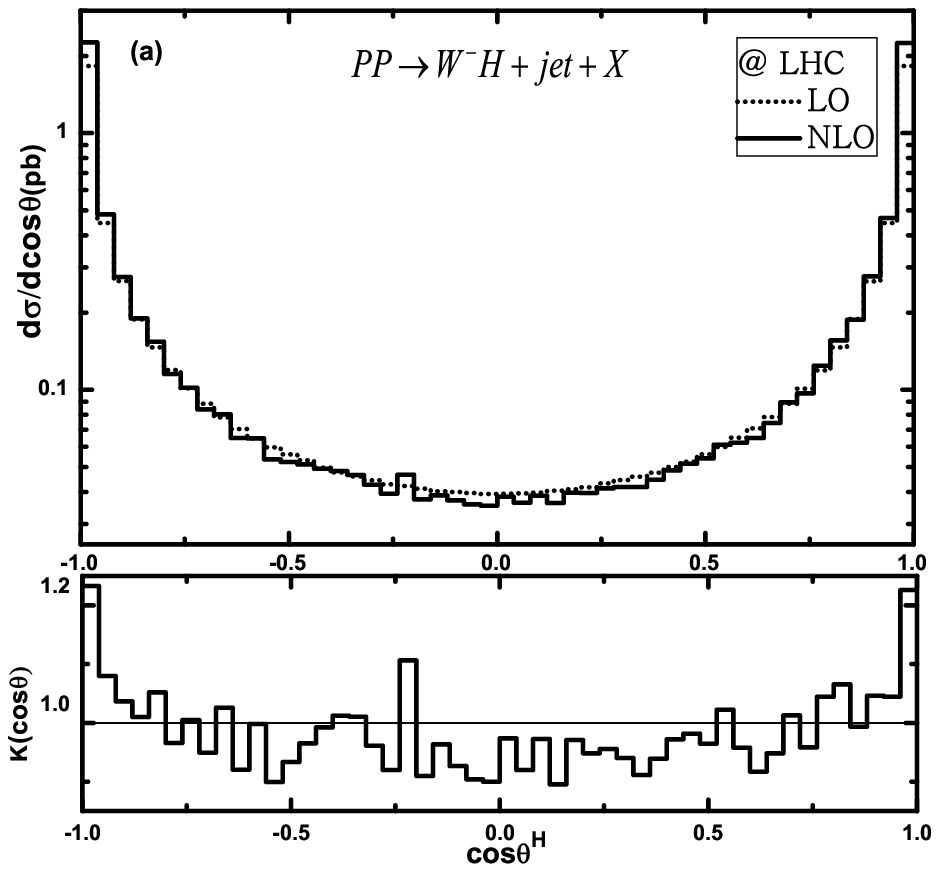}
\includegraphics[scale=0.65]{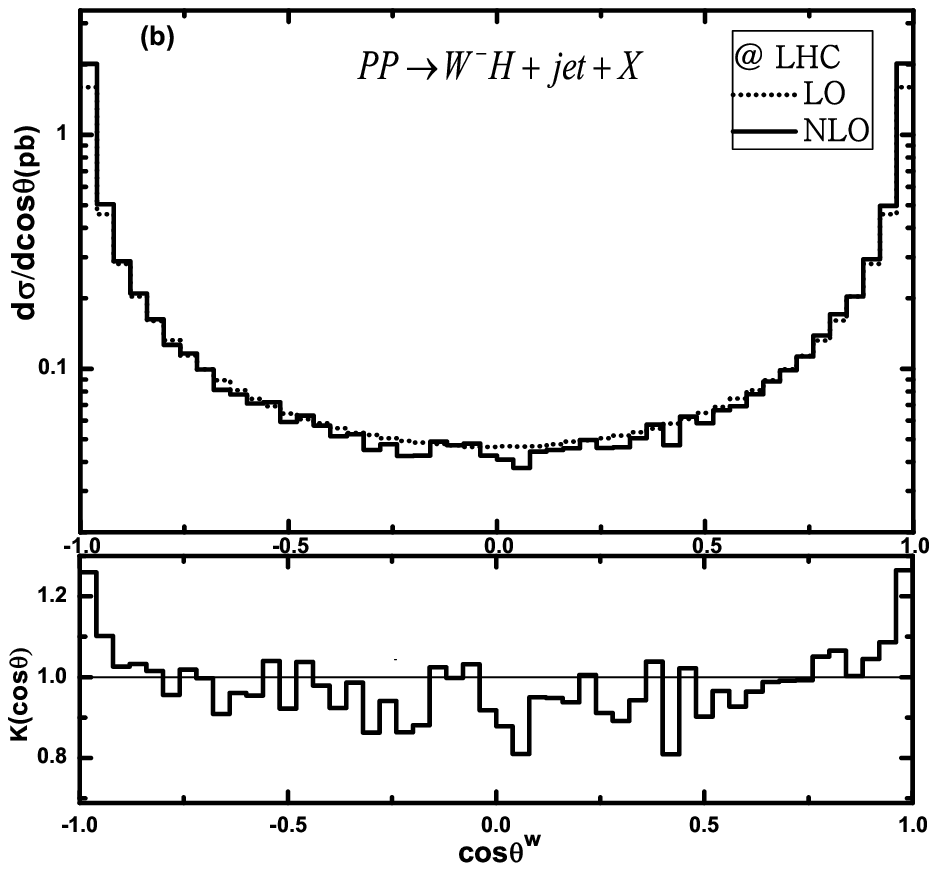}
\includegraphics[scale=0.65]{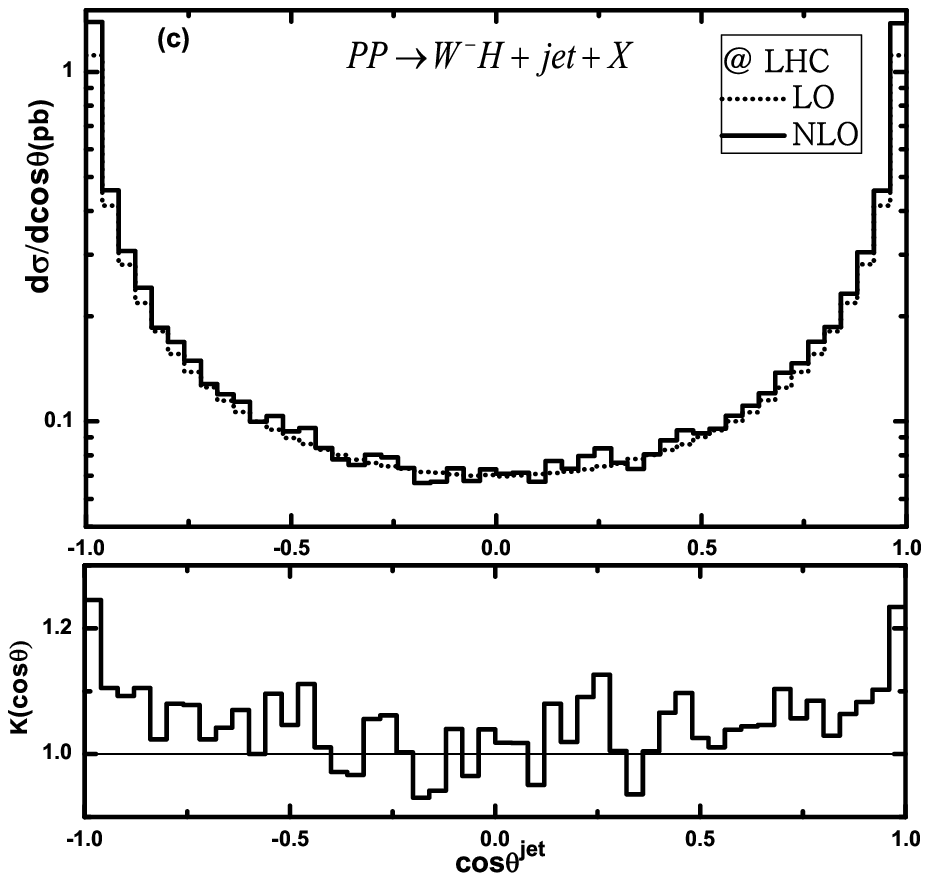}
\caption{\label{fig10} The LO, NLO QCD corrected
distributions($\frac{d\sigma}{d\cos\theta}$) and the corresponding
K-factors($K(\cos\theta)\equiv\frac{d\sigma_{NLO}}
{d\cos\theta}/\frac{d\sigma_{LO}} {d\cos\theta}$) versus the
cosine of the angle between the final particle and the direction
of one of the incoming protons for the process \ppwhjm at the LHC
by applying the inclusive scheme with $p_{T,j}^{cut}=20~GeV$ and
taking $m_H=120~GeV$ and $\mu=\mu_0$. (a) for final Higgs boson,
(b) for final $W^-$-boson, (c) for final leading jet. }
\end{figure}
\begin{figure}
\centering
\includegraphics[scale=0.65]{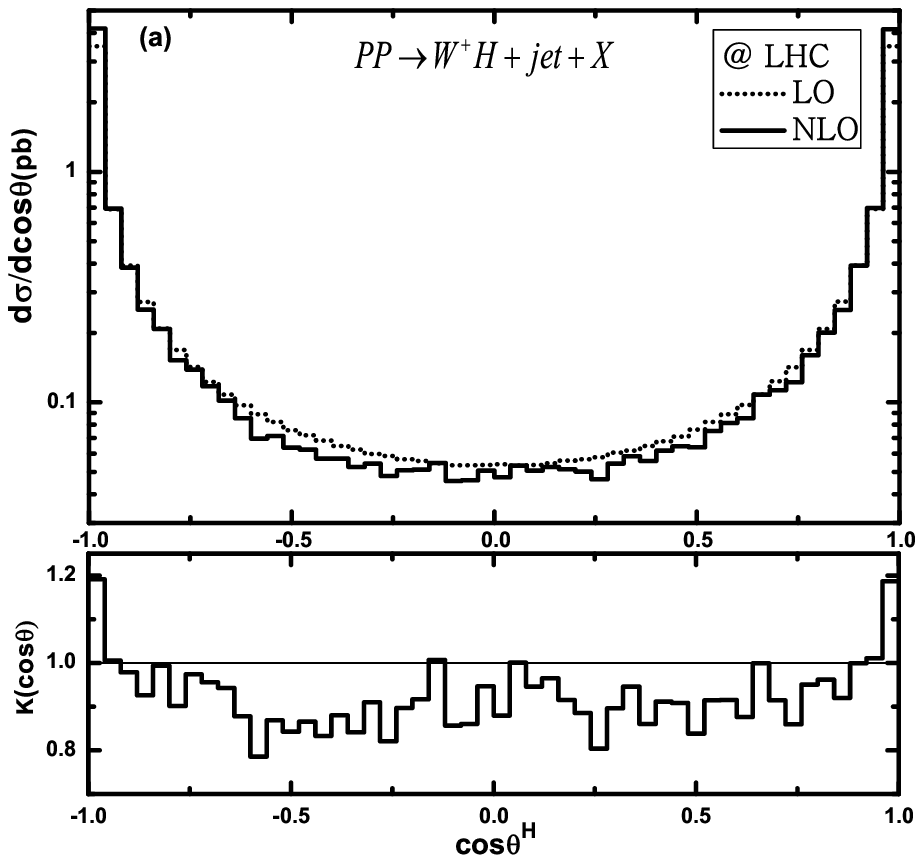}
\includegraphics[scale=0.65]{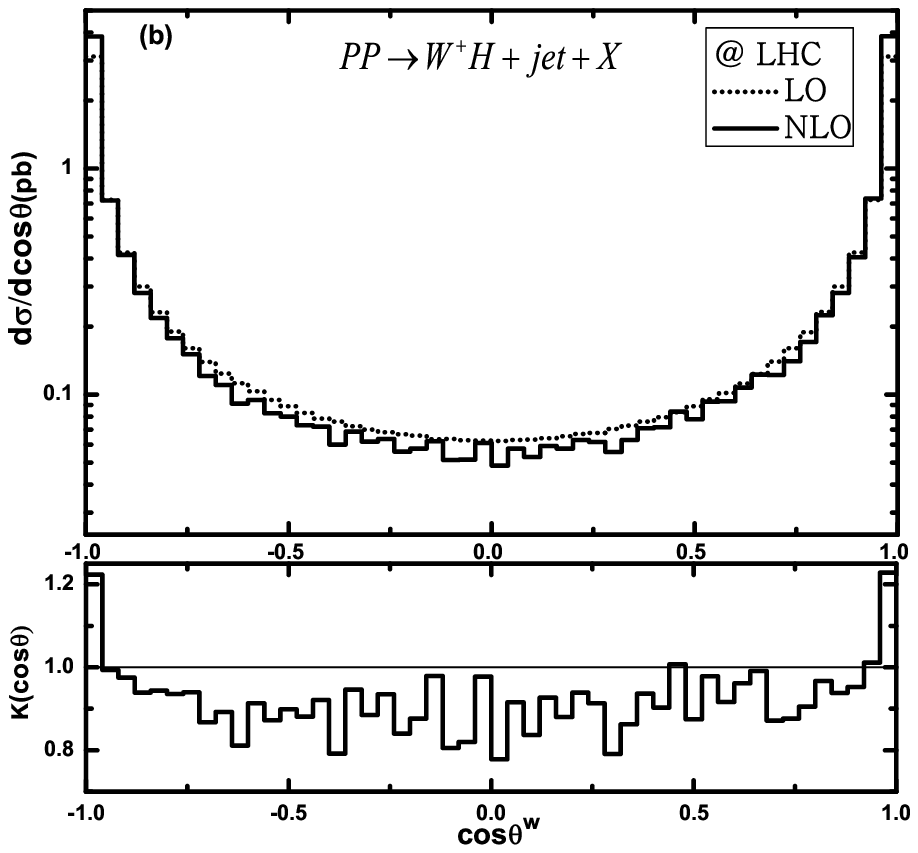}
\includegraphics[scale=0.65]{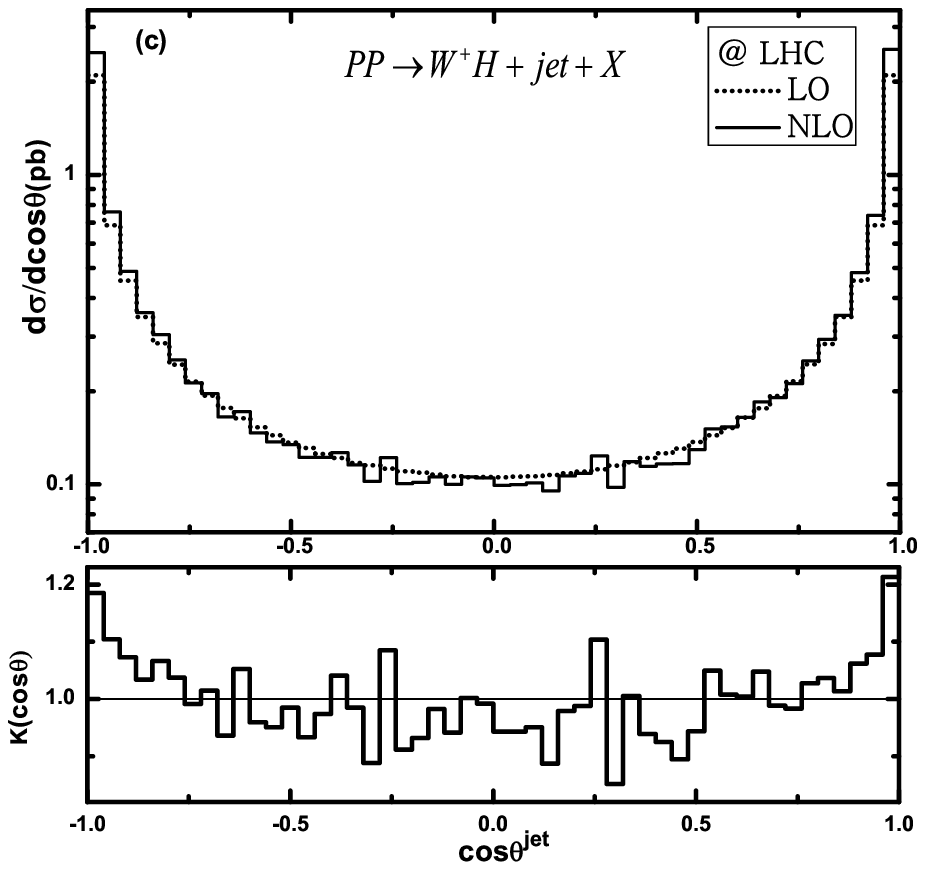}
\caption{\label{fig11} The LO, NLO QCD corrected
distributions($\frac{d\sigma}{d\cos\theta}$) and their
corresponding K-factor($K(\cos\theta)\equiv\frac{d\sigma_{NLO}}
{d\cos\theta}/\frac{d\sigma_{LO}} {d\cos\theta}$) versus the
cosine of the angle of the final particle with respect to the direction
of one of the incoming protons for the process \ppwhjp at the LHC
by applying the inclusive scheme with $p_{T,j}^{cut}=20~GeV$ and
taking $m_H=120~GeV$ and $\mu=\mu_0$. (a) for final Higgs boson,
(b) for final $W^+$-boson, (c) for final leading jet. }
\end{figure}

We take the orientation of the incoming antiproton as the z-axis
direction at the Tevatron and the orientation of one of the
incoming protons as the direction of z-axis at the LHC . The
$\theta^{H}$($\theta^{W}$ or $\theta^{j}$) is define as the
$H$-boson ($W$-boson or the leading jet) production angle with
respect to the z-axis direction for the $W^{\pm}H^0j$ production
process at the Tevatron or the LHC. In Figs.\ref{fig9}(a,b,c), we
present the LO and NLO QCD corrected distributions as the
functions of the cosines of the $H^0$-boson, $W^-$-boson and
leading jet production angles($\frac{d\sigma}{d\cos\theta}$), and
their corresponding
K-factors($K(\cos\theta)\equiv\frac{d\sigma_{NLO}}
{d\cos\theta}/\frac{d\sigma_{LO}}{d\cos\theta}$) at the Tevatron.
The NLO distributions in Figs.\ref{fig9}(a,b,c) are obtained by
applying the inclusive scheme with $p_{T,j}^{cut}=20~GeV$, and taking
$m_H=120~GeV$, $\mu=\mu_0$. They show that the produced
$H^0$-boson, $W^-$-boson and leading jet slightly prefer to go out
in the forward hemisphere region at the Tevatron. In
Figs.\ref{fig10}(a,b,c) and Figs.\ref{fig11}(a,b,c), we present
the LO and NLO QCD corrected differential cross sections and their
corresponding K-factors as the functions of the cosines of the
$H^0$-boson, $W^-$(or $W^{+}$)-boson and leading jet production
angles for the process $pp \to W^{\pm}Hj+X$ at the LHC,
separately. Again in these figures we apply the inclusive scheme
with $p_{T,j}^{cut}=20~GeV$, denote $p_T^{j}$ as the transverse
momentum of the leading jet and set $m_H=120~GeV$, $\mu=\mu_0$.
Both the LO and NLO curves in Figs.\ref{fig10}(a,b,c) and
Figs.\ref{fig11}(a,b,c) demonstrate that the outgoing $H^0$-boson
$W^-$-boson and leading jet are symmetrically distributed in the
forward and backward hemisphere regions.

\section{Summary}
\par
In this paper we calculate the full NLO QCD corrections to the
$W^{\pm}H^0$ production associated with a jet at the Tevatron Run
II and the LHC. We investigate the dependence of the integrated
cross sections on the energy scale, and study the NLO QCD
contributions to the differential cross sections of the transverse
momenta($\frac{d\sigma}{dp_T}$) and the production angle
distributions($\frac{d\sigma}{d\cos\theta}$) for the final
particles at both hadronic colliders. We find that the NLO QCD
radiative corrections obviously modify the LO integrated and
differential cross sections, and the NLO QCD corrections to the
$W^{\pm}H^0$+jet production processes significantly reduce the
scale uncertainties of the LO cross sections at both hadron
colliders. Our numerical results show that in conditions of
applying the inclusive scheme with $p_{T,j}^{cut}=20~GeV$, taking
$\mu=\mu_0$ and $m_H=120~GeV$, the K-factor for the process $p\bar
p \to W^{\pm}H^0j+X$ at the Tevatron Run II is $1.15$, while the
K-factors for the $pp \to W^-H^0j+X$ and $pp \to W^+H^0j+X$
processes at the LHC are $1.12$ and $1.08$, respectively. We
conclude that in studying the hadronic $WH^0$ production channel
the NLO QCD corrections to the $WH^0j$ production process which is
part of inclusive $WH^0$ production, should be taken into account.

\vskip 5mm
\par
\noindent{\large\bf Acknowledgments:} This work was supported in
part by the National Natural Science Foundation of
China(No.10875112), the Specialized Research Fund for the Doctoral
Program of Higher Education(SRFDP)(No.20093402110030), the
National Science Foundation for Post-doctoral Scientists of China
(No.20080440103), and the Funds for Creative Research Program of
USTC.

\end{document}